\documentclass[preprintnumbers,prd,onecolumn,floatfix,superscriptaddress,nofootinbib]{revtex4}
\usepackage{graphicx}
\usepackage{epsfig}
\usepackage{bm}
\usepackage{amssymb}
\usepackage{float}
\usepackage{amsmath}
\usepackage{subfigure}
\usepackage{dcolumn}
\usepackage{cancel}
\usepackage[colorlinks]{hyperref}
\usepackage[usenames,dvipsnames]{color}
\usepackage{eurosym}
\usepackage{amsfonts}
\usepackage{graphicx,epsfig}
\usepackage{epstopdf}
\usepackage{amssymb}

\hypersetup{
breaklinks=true,pdfstartview={FitH}, colorlinks=true, linkcolor=blue, citecolor=red, filecolor=magenta,    
urlcolor=blue, anchorcolor=green, linktocpage=true}

\def\doi{http://doi.org}

\begin{document}

\title{An Accelerating Cosmological Model from a Parametrization of Hubble
Parameter}
\author{S. K. J. Pacif}
\email{shibesh.math@gmail.com}
\affiliation{Department of Mathematics, School of Advanced Sciences, VIT University,
Vellore 632014, India}
\author{Md Salahuddin Khan}
\email{salahuddinnic.jmi@gmail.com}
\affiliation{Centre for Theoretical Physics, Jamia Millia Islamia, New Delhi 110025, India}
\author{L. K. Paikroy}
\email{lplpaikray@gmail.com}
\affiliation{Department of Mathematics, Dalmia College, Raj Gangpur 770017, India}
\author{Shalini Singh}
\email{shalinisingh@jssaten.ac.in}
\affiliation{Department of Applied Mathematics, JSS Academy of Technical Education, Noida
201301, India}

\begin{abstract}
In view of late-time cosmic acceleration, a dark energy cosmological model
is revisited wherein Einstein's cosmological constant is considered as a
candidate of dark energy. Exact solution of Einstein field equations (EFEs)
is derived in a homogeneous isotropic background in classical general
relativity. The solution procedure\ is adopted, in a model independent way
(or the cosmological parametrization). A simple parametrization of the
Hubble parameter ($H$) as a function of cosmic time `$t$' is considered
which produces an exponential type of evolution of the scale factor ($a$)
and also shows a negative value of deceleration parameter at the present
time with a signature flip from early deceleration to late acceleration.
Cosmological dynamics of the model obtained have been discussed
illustratively for different phases of the evolution of the universe. The
evolution of different cosmological parameters are shown graphically for
flat and closed cases of Friedmann-Lemaitre-Robertson-Walker (FLRW)
space-time for the presented model (open case is incompatible to the present
scenario). We have also constrained our model parameters with the updated ($%
36$ points) observational Hubble dataset.
\end{abstract}

\maketitle


\section{Introduction}

\qquad One of the fundamental problem in Standard big bang cosmology (SBBC)
is the long-standing cosmological constant problem \cite{CCP1, CCP2} and
there were several attempts to solve this problem in the late eighties and
nineties \cite{LM1, LM2, LM3, LM4, LM5, LM6, LM7}. Another problem is the
cosmic age problem \cite{ageP1, ageP2} i.e., some objects in the universe
were estimated to be older than the time elapsed since the Big Bang. An
intriguing problem came to exist in 1998 with the observations of supernovae
of type Ia which provided the results against the all-time decelerating
expansion of SBBC due to dominance of gravitational pull \cite{HZ1, SCP1}.
The observations brought the new concept of late-time \textit{cosmic
acceleration}. Now, cosmic acceleration has become a very important issue to
be discussed in frontline cosmology and is getting supported by some more
robust observations \cite{CMB1, CMB2, BAO1, BAO2, SDSS1, SDSS2, SDSS3}. The
idea of late-time acceleration yields several new modifications in the
general theory of relativity (GTR). One such modification is the inclusion
of a new form of energy known as \textit{Dark Energy} (DE) into the Einstein
field equations. Now, the theory of DE has taken a special status in the
contemporary cosmology. The theory resolves not only the all time
decelerated expansion problem of standard cosmology but also the age
problem. In the past twenty years, several surveys have been done at
theoretical as well as observational ground on dark energy \cite{DES1, DES2,
Plank1, Plank2}. The current observations reveals that the universe is
estimated to have $95\%$ filled with dark matter and dark energy and only $%
5\%$ of baryonic matter. We don't know much about this mysterious DE, but it
is assumed to be homogeneous and permeates all over the space and must
possess high negative pressure that is responsible for the cosmic speed up.
Moreover, there are strong debates on the candidature of the dark energy.
The simplest one being the Einstein's cosmological constant $\Lambda $ which
on adding to the EFEs in a homogeneous isotropic FLRW background (known as $%
\Lambda $CDM model \cite{CC1, CC2, CC3, CC4}) is still best suited to many
cosmological observations.

\qquad The presented work is an attempt to address late-time cosmic
acceleration in an FLRW background. Einstein field equations in FLRW
background contains two independent equations with three unknowns (energy
density- $\rho $, pressure- $p$ and scale factor- $a$) which can be solved
by assuming the equation of state. With the addition of an extra degree of
freedom - \textit{dark energy,} the system becomes undeterminable. There
exists a number of ways to deal with this inconsistency in literature. We,
here adopt a very simple mathematical approach to find the exact solution of
the field equations known as model independent way or cosmological
parametrization. An exciting feature of the functional form of $H$
considered here is that the deceleration parameter $q$ shows signature flip
that describes a universe from early deceleration to present acceleration as
expected by observations. The dynamics of the universe in different phases
of evolution is discussed here for two different i.e. flat and closed cases
of FLRW space-time. The open case is incompatible with the present scenario.
The work is organized as follows: Sect. 1 provides a brief introduction to
some problems of GTR and dark energy. In Sect. 2, we review the derivation
of the field equations and discuss the solution technique. Also, we discuss
the geometrical interpretation of the obtained model in Sect. 2. In Sect. 3,
we discuss the dynamics of the obtained model and analyze the behavior of
the physical parameters and describe the evolution in the radiation
dominated and matter dominated era of the universe. Also, we show the
evolution of cosmological parameters through graphical representation in
Sect. 3. Finally, we conclude with our results in Sect. 4.

\section{Basic equations and solution of field equations}

\subsection{Field equations in GR}

\qquad We consider a homogeneous and isotropic Robertson-Walker space-time
given by the equation, 
\begin{equation}
ds^{2}=-c^{2}dt^{2}+a^{2}(t)\left[ \frac{dr^{2}}{1-kr^{2}}+r^{2}d\Omega ^{2}%
\right] .  \label{FRWEQ}
\end{equation}%
with $8\pi G=c=1$. The matter source in the universe is provided by the
total energy-momentum tensor (EMT) given by the equation, 
\begin{equation}
T_{\mu \nu }^{Total}=(\rho _{Total}+p_{Total})u_{\mu }u_{\nu
}+p_{Total}~g_{\mu \nu },  \label{1}
\end{equation}%
where $T_{\mu \nu }^{Total}$ is the EMT for the two energy components in the
universe i.e. $\rho _{Total}=\rho +\rho _{de}$, where $\rho =\rho _{r}+\rho
_{m}$ and $p_{Total}=p+p_{de}$, where $p=p_{r}+p_{m}$ are the energy
densities and pressures for each component. Here and afterwards the suffix `$%
r$' and `$m$' stands for the radiation and matter components respectively
for the corresponding quantity and suffix `$de$' stands for the dark energy.

The Einstein field equation with total energy momentum tensor is, 
\begin{equation}
R_{\mu \nu }-\frac{1}{2}Rg_{\mu \nu }=-T_{\mu \nu }^{Total}\text{ \ \ \
(with }8\pi G=1\text{),}  \label{2}
\end{equation}%
yield two independent equations as follows,%
\begin{equation}
\rho _{Total}=3\frac{\dot{a}^{2}}{a^{2}}+3\frac{k}{a^{2}},  \label{3}
\end{equation}%
\begin{equation}
p_{Total}=-2\frac{\ddot{a}}{a}-\frac{\dot{a}^{2}}{a^{2}}-\frac{k}{a^{2}},
\label{4}
\end{equation}%
where an overhead dot ($\cdot $)\ represents ordinary derivative with
respect to cosmic time `$t$' only. We believe that the interaction between
the two matter components are natural. From equations (\ref{3}) and (\ref{4}%
), one can easily derive the equation of continuity as

\begin{equation}
\dot{\rho}_{Total}+3\frac{\dot{a}}{a}(\rho _{Total}+p_{Total})=0.
\label{contin}
\end{equation}
We can see that from equations (\ref{3}) and (\ref{4}) and (\ref{contin}),
there are only two independent equations in five variables $a$, $\rho $, $p$%
, $\rho _{de}$, $p_{de}$.

\qquad We consider the usual barotropic equation of state for normal
(/ordinary) matter 
\begin{equation}
p=w\rho ,  \label{eos1}
\end{equation}%
where, $w=\frac{1}{3}$ for radiation component and $w=0$ for pressure-less
dust component in the universe.\newline

\qquad We now solve these equations to discuss the cosmic history for
different phases of evolution separately i.e., in the early radiation
dominated (RD) era following the late matter dominated (MD) era.

\subsection{Parametrization of H}

\qquad In literature, there are many physical arguments and motivations on
the model independent way to study the dynamics of dark energy models \cite%
{LINDER1, LINDER2}. In this section, we follow the same idea of cosmological
parametrization and solve the field equations explicitly and also discuss
the dynamics of the universe in different phases of evolution of the
universe. In order to describe certain phenomena of the universe e.g.,
cosmological phase transition from early inflation to deceleration and
deceleration to late time acceleration, many theoreticians have considered
different parametrization of cosmological parameters, where the model
parameters involved in the parametrization can be constrained through
observational data. Most of the parametrization deal with the equation of
state parameter $w(z)$ \cite{LINDER3} or deceleration parameter $q(z)$ \cite%
{CHUNA}. Some well known parametrization are Chevelier-Porrati-Linder (CPL)
parametrization \cite{CPL1}, Jassal-Bagla-Padmanabhan parametrization \cite%
{JBL1} on $w(z)$. A critical review of this argument shows, one can
parametrize other geometrical or physical parameters also. Pacif et al. \cite%
{PACIF1} have summarized these parametrization of the physical and
geometrical parameters in some detail and also proposed a new
parametrization on Hubble parameter. Here, we consider the parametrization
of the Hubble parameter in the form \cite{JPS, BAN} 
\begin{equation}
H(a)=\alpha (1+a^{-n}),  \label{H1}
\end{equation}%
where $\alpha >0$ and $n>1$ are constants (better call them model
parameters).

Equation (\ref{H1}) readily give the explicit form of scale factor as, 
\begin{equation}
a(t)=(ce^{n\alpha t}-1)^{\frac{1}{n}},  \label{sf1}
\end{equation}%
where, $c\neq 0$ is a constant of integration. We can see that the form of
scale factor is an exponential function and contain two model parameters $n$
and $\alpha $\ which regulate the dynamics of the evolution. As $%
t\rightarrow 0$, we can have $a(0)=(c-1)^{\frac{1}{n}}=a^{(i)}$ (say, the
superscript $(i)$ stands for initial value of the parameter as $t\rightarrow
0$), which is non-zero for $c\neq 1$ implying a non-zero initial value of
the scale factor (or a cold initiation of the universe with a finite volume).

\subsection{Geometrical interpretation of the model}

The first and second time-derivatives of the functional form of the scale
factor (\ref{sf1}) are given by 
\begin{equation}
\dot{a}=\alpha ce^{n\alpha t}\left[ ce^{n\alpha t}-1\right] ^{\frac{1}{n}-1},
\label{18}
\end{equation}%
and 
\begin{equation}
\ddot{a}=\alpha ^{2}ce^{n\alpha t}\left[ ce^{n\alpha t}-n\right] \left[
ce^{n\alpha t}-1\right] ^{\frac{1}{n}-2}.  \label{19}
\end{equation}

\qquad These indicate, in the beginning, initially at time $t=0$, the
velocity and the acceleration of the universe are $\dot{a}^{(i)}=\alpha
c(c-1)^{\frac{1}{n}-1}$ and $\ddot{a}^{(i)}=\alpha ^{2}c(c-n)(c-1)^{\frac{1}{%
n}-2}$ which depict that the obtained model starts with a finite volume, a
finite velocity and a finite acceleration. This is a notable deviation from
the standard model. The expressions for the Hubble parameter and
deceleration parameter in terms of cosmic time `$t$' is written with the
help of equation (\ref{sf1}) as

\begin{equation}
H(t)=\frac{\alpha ce^{n\alpha t}}{ce^{n\alpha t}-1}.  \label{hp}
\end{equation}

and

\begin{equation}
q(t)=-1+\frac{n}{ce^{n\alpha t}}.  \label{20}
\end{equation}%
Equation (\ref{20}) shows that the deceleration parameter is time dependent
which can take both positive and negative value i.e. at the initial stage of
the evolution, for small $t$, the second term in (\ref{20}) will dominate
over first term if $n>1$ and $q$ will be positive and at later stage of the
evolution, for large $t$ (as $t\rightarrow \infty $), the second term will
be small and effectively zero and $q$ approach to $-1$. In fact, we can see,
from equation (\ref{20}) as $t\rightarrow 0$, $q=-1+\frac{n}{c}$ which is a
constant and is positive for $n>1$ and $c<n$. This imply that the
deceleration parameter has a signature flip from positive to negative with
the evolution. So, this model exhibits a early deceleration to late
acceleration which is suitable for structure formation in the early stage of
evolution and accelerated expansion in the later stage of the evolution.
From equation (\ref{hp}), we can observe that as $t\rightarrow 0$, $H^{(i)}=%
\frac{\alpha c}{c-1}>0$ for $c>1$, which is constant. Also, $H(t)$ is a
decreasing function of time and attains a constant value $`\alpha $' as $%
t\rightarrow \infty $. To have a rough sketch of the evolution of the
geometrical parameters ($a$, $H$, $q$) of the model, we shall choose the
integrating constant $c$ and model parameters $\alpha $ and $n$ in such a
way that the evolution of the cosmological parameters could be in accordance
with the observations. By some analytical choice, we have chosen $c=1.2$, $%
n=1.43$ and $\alpha =0.1$ (and two more values of $n$ and $\alpha $ in the
neighborhood $i.e.$ $n=1.23$, $1.43$, $1.63$ and $\alpha =0.08$, $0.10$, $%
0.12$, just to have a better understanding of the effect of the model
parameters on the evolution) arbitrarily and with suitable time units. The
early evolution of $a(t)$, $H(t)$ and $q(t)$ are shown graphically with
these values in the following FIG.\ref{fig:1} \& FIG.\ref{fig:2} as an
exemplification.

\begin{figure}[tbph]
\begin{center}
{\scriptsize $%
\begin{array}{c@{\hspace{.1in}}ccc}
\includegraphics[width=2in]{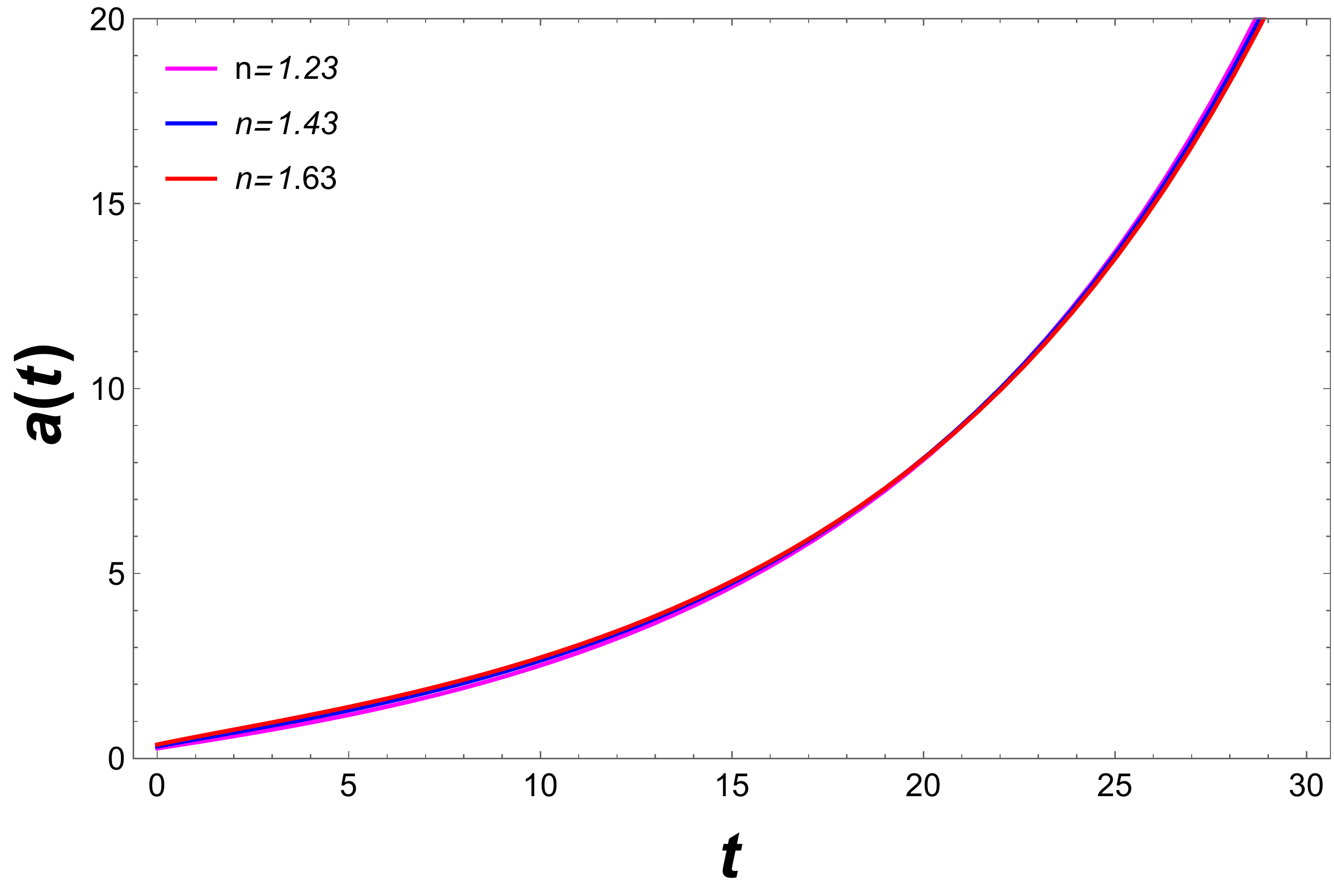} & \includegraphics[width=2in]{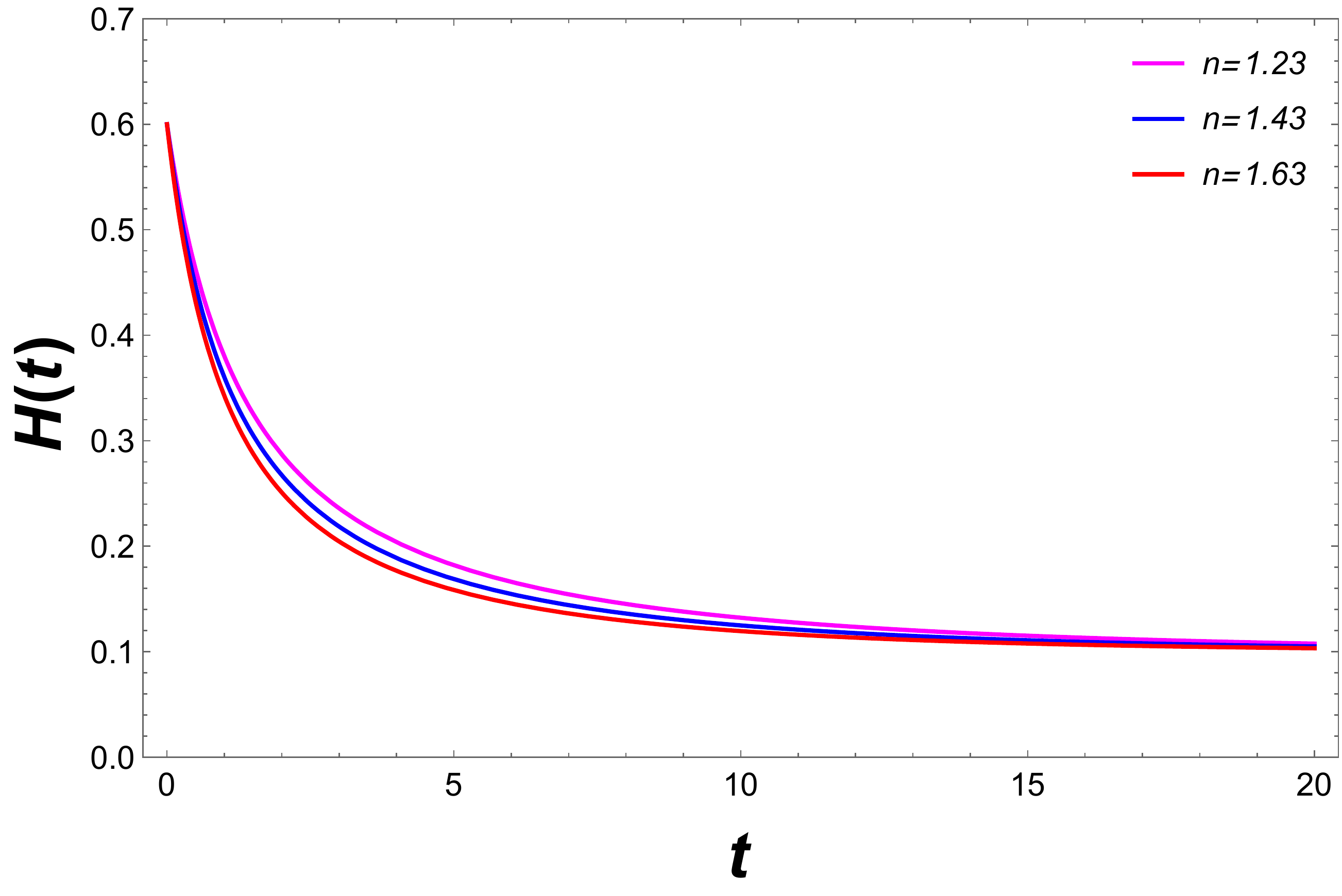} & %
\includegraphics[width=2in]{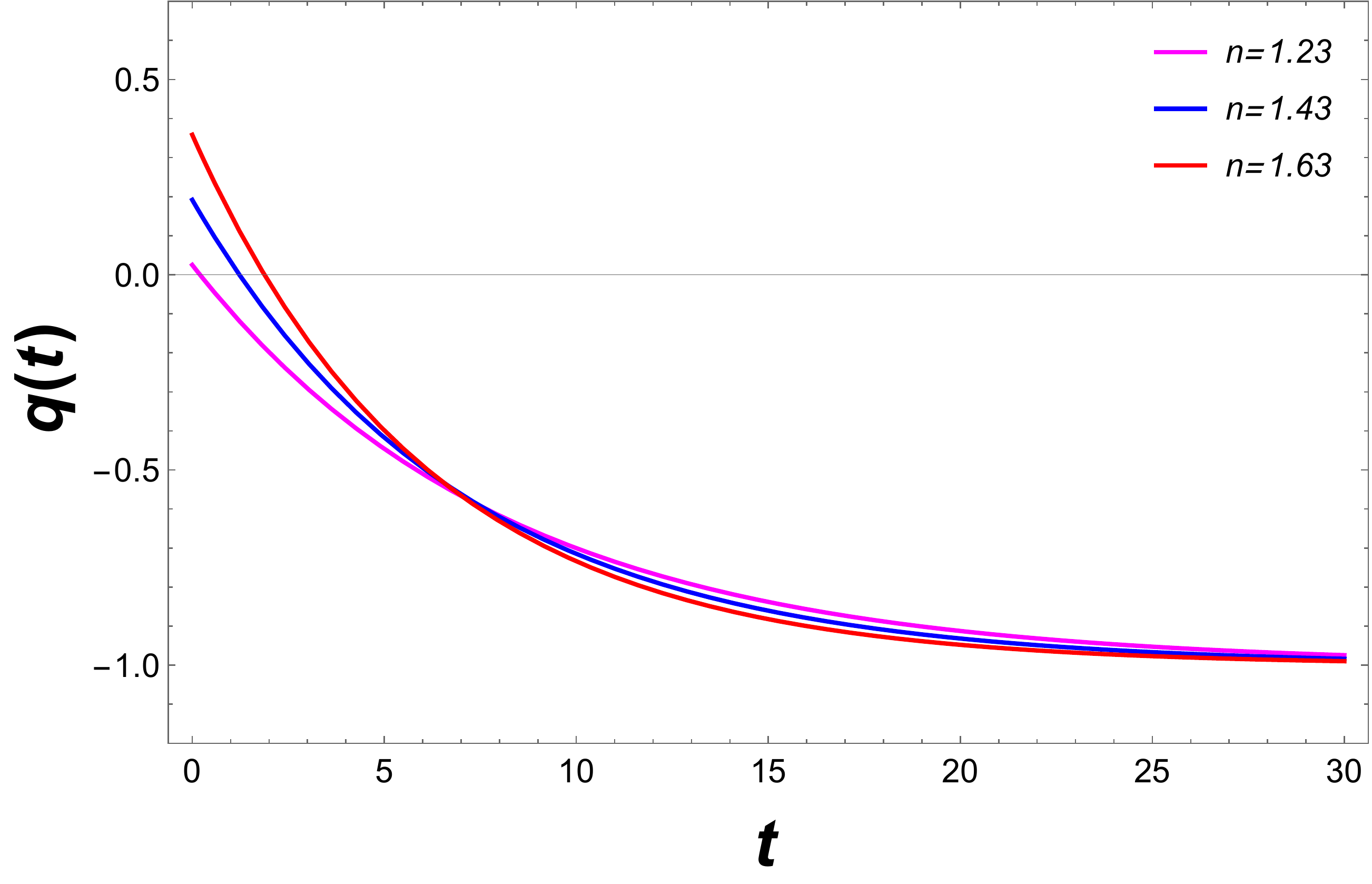} &  \\ 
\mbox (a) & \mbox (b) & \mbox (c) & 
\end{array}%
$ }
\end{center}
\caption{ The plot shows a sketch of the time evolution of the geometrical
parameters (a) Scale factor `$a$' (b) Hubble parameter `$H$' (c)
Deceleration parameter `$q$' with suitable units of cosmic time `$t$'. For
these plots, we have chosen the integrating constant $c=1.2$, the model
parameter $\protect\alpha =0.1$ is fixed with different values of $%
n=1.23,1.43,1.63$.}
\label{fig:1}
\end{figure}

\begin{figure}[tph]
\begin{center}
{\scriptsize $%
\begin{array}{c@{\hspace{.1in}}ccc}
\includegraphics[width=2in]{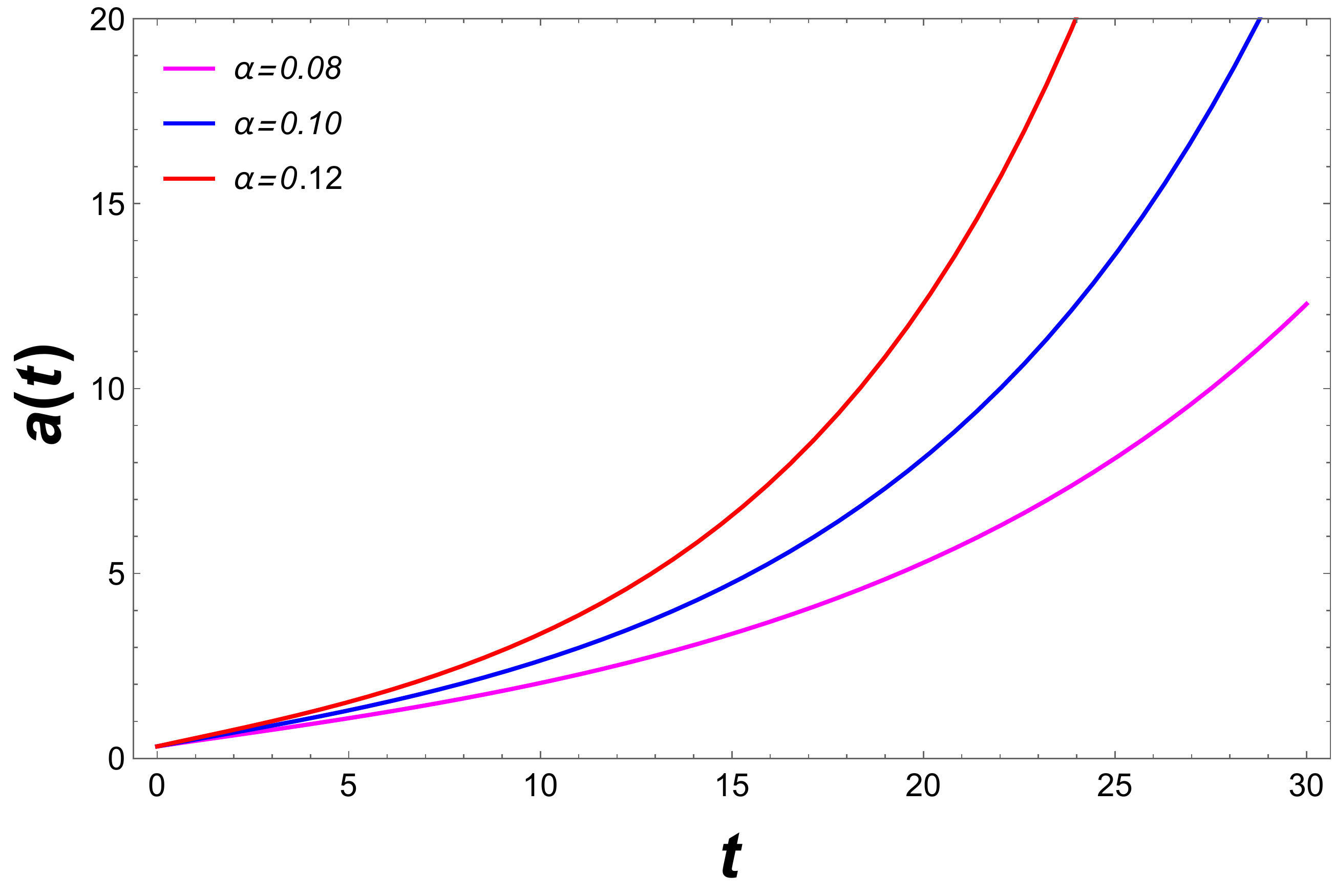} & \includegraphics[width=2in]{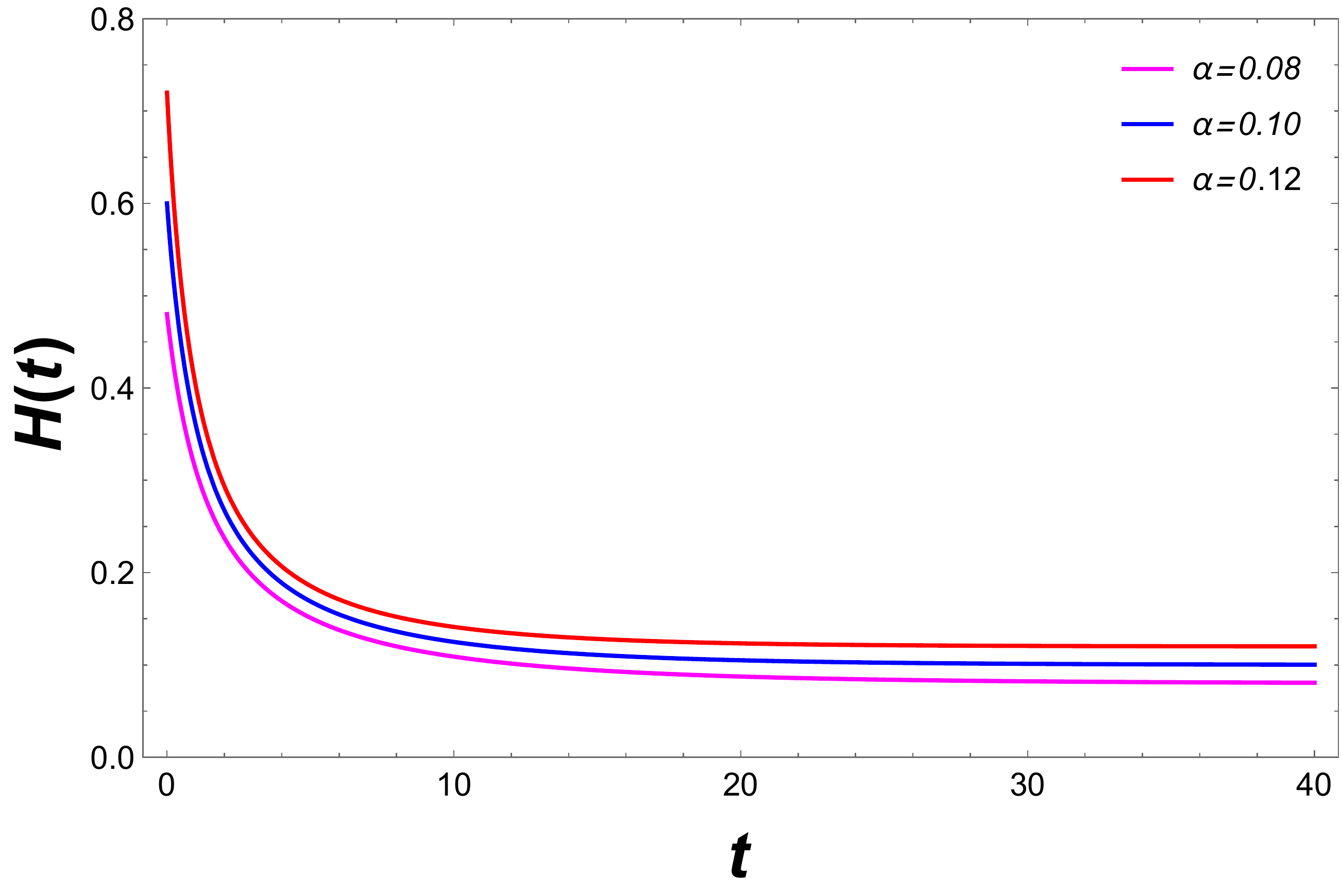} & %
\includegraphics[width=2in]{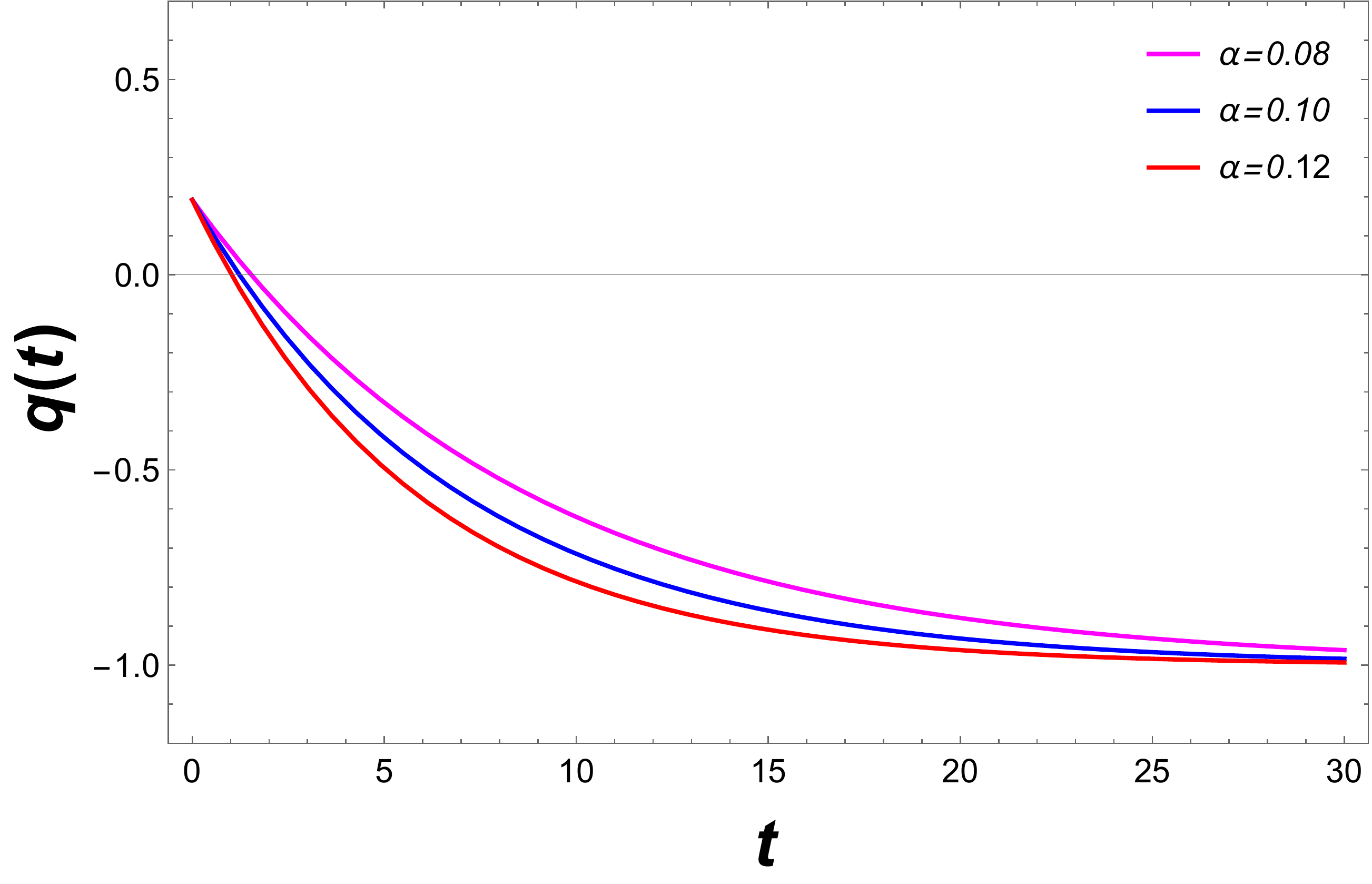} &  \\ 
\mbox (a) & \mbox (b) & \mbox (c) & 
\end{array}%
$ }
\end{center}
\caption{The plot shows a sketch of the time evolution of the geometrical
parameters (a) Scale factor `$a$' (b) Hubble parameter `$H$' (c)
Deceleration parameter `$q$' with suitable units of cosmic time `$t$'. For
these plots, we have chosen the integrating constant $c=1.2$, the model
parameter $n=1.43$ is fixed with different values of $\protect\alpha %
=0.08,0.10,0.12$.}
\label{fig:2}
\end{figure}

\section{Dynamics and physical interpretation of the model}

\qquad Equations (\ref{3}) and (\ref{4}) with the help of (\ref{eos1}) can
be rewritten as,%
\begin{equation}
\rho +\rho _{de}=3H^{2}+3\frac{k}{a^{2}},  \label{new1}
\end{equation}%
\begin{equation}
w\rho +p_{de}=\left( 2q-1\right) H^{2}-\frac{k}{a^{2}}.  \label{new2}
\end{equation}

\qquad We can observe that the right hand side of the above system of
equations are known functions of cosmic time $t$ with the time-dependent
functions of $a,$ $q,$ $H$ given in (\ref{sf1}), (\ref{20}), (\ref{hp}). In
the left hand side, we have three unknowns functions $\rho ,$ $\rho _{de},$ $%
p_{de}$. The general equation of state of dark energy can be represented as, 
\begin{equation}
w_{de}=\frac{p_{de}}{\rho _{de}}.  \label{new3}
\end{equation}

\qquad The parameter $w_{de}$ may be a constant or in general, a
time-dependent function that evolves with the evolution of the universe. The
time-dependence of $w_{de}$ results in a plethora of dark energy
cosmological models of the universe. There is not much idea about the
candidate of dark energy for which, the equation of state of dark energy $%
w_{de}$ is unknown. For scalar field models, astrophysical data indicate the
effective equation of state parameter $w_{eff}\left( =\frac{p_{tot}}{\rho
_{tot}}=\frac{p+p_{de}}{\rho +\rho _{de}}\right) $ lies in the interval $%
-1.48<w_{eff}<-0.72$ \cite{EoS1a}-\cite{EoS2}. The analysis of the
observational data mildly favor models of dark energy with $w_{eff}$
crossing $-1$ line in the recent past \cite{EoS2}, \cite{linecros}. For
detailed reviews on dark energy and the candidates of dark energy, see \cite%
{DEREV1, DEREV2, DEREV3, DEREV4}. However, Einstein's cosmological constant
is a favorable candidate for dark energy ($\Lambda $CDM model compatible
with observations) for which $w_{de}$ take a constant value $-1$. So, we
consider the cosmological constant (CC) as a candidate of dark energy and
continue our analysis further. For the CC, the equation of state becomes, $%
p_{de}=-\rho _{de}$. In this case equations (\ref{new1}) and (\ref{new2})
can be solved to give the expressions for the matter ($+$ radiation) energy
density $\rho $ and dark energy density as\textbf{\ } 
\begin{eqnarray}
\rho &=&\frac{2}{1+w}\left[ \left( 1+q\right) H^{2}+\frac{k}{a^{2}}\right] ,
\label{new4} \\
\rho _{de} &=&\frac{1}{1+w}\left[ \left( 1+3w-2q\right) H^{2}+\left(
1+3w\right) \frac{k}{a^{2}}\right] .  \label{new5}
\end{eqnarray}

\qquad Now, we can discuss the dynamics of the obtained model in different
phases of evolution of the universe for three different cases in FLRW
geometry i.e. flat ($k=0$), closed ($k=1$) and open ($k=-1$).

\subsection{Radiation Dominated universe}

\qquad In the early \textit{pure radiation era}, we have $w=\frac{1}{3}$ and 
$\rho \approx \rho _{r}$. Equations (\ref{new4}) and (\ref{new5}) together
with (\ref{sf1}) yield the expressions for the radiation energy density and
the dark energy density in the early phase of evolution and are given by, 
\begin{eqnarray}
\rho _{r} &=&\frac{3}{2}\left[ \frac{n\alpha ^{2}ce^{n\alpha t}}{\left[
ce^{n\alpha t}-1\right] ^{2}}+\frac{k}{\left[ ce^{n\alpha t}-1\right] ^{%
\frac{2}{n}}}\right] ,  \label{new6} \\
\rho _{de} &=&\frac{3}{2}\left[ \frac{\alpha ^{2}(2ce^{n\alpha
t}-n)ce^{n\alpha t}}{\left[ ce^{n\alpha t}-1\right] ^{2}}+\frac{k}{\left[
ce^{n\alpha t}-1\right] ^{\frac{2}{n}}}\right] .  \label{new7}
\end{eqnarray}

\qquad The very early universe (from the beginning to a pico second after in
SBBC) is generally addressed to quantum gravity and a full theory of quantum
gravity is not available till. So, we keep ourselves at the classical level
only for our analysis. Equations (\ref{new6}) and (\ref{new7}) provide the
evolution of the energy densities in the radiation dominated era (i.e. after 
$10^{-32}$ to $4.7\times 10^{4}$ years in SBBC) and the expressions (\ref%
{new6}) and (\ref{new7}) can not be valid near the Plank epoch. However, at
classical level, we may take these expressions for consideration. As $%
t\rightarrow 0$, we can have $\rho _{r}^{(i)}\approx 1.5\times \left[
n\alpha ^{2}c(c-1)^{-2}+k(c-1)^{-\frac{2}{n}}\right] $ and $\rho
_{de}^{(i)}\approx 1.5\times \left[ \alpha ^{2}(2c-n)c(c-1)^{-2}+k(c-1)^{-%
\frac{2}{n}}\right] $ suggesting that $\rho _{r}^{\left( i\right) }>0$ in
the beginning provided $c\neq 1$ and $\rho _{de}^{(i)}>0$ provided $n<2c$
and $c\neq 1$ for flat and closed universe. We can observe, from equations (%
\ref{new6}) and (\ref{new7}) that the positivity condition for $\rho $ and $%
\rho _{de}$ holds good for the above choice of $n$, $c$, $\alpha $ in the
cases of flat ($k=0$) and closed ($k=1$) geometry. But, for open ($k=-1$)
geometry, the second term in equations (\ref{new6}) and (\ref{new7}) must
not dominate over the first term i.e. $\frac{n\alpha ^{2}ce^{n\alpha t}}{%
\left[ ce^{n\alpha t}-1\right] ^{2}}>\frac{1}{\left[ ce^{n\alpha t}-1\right]
^{\frac{2}{n}}}$ or $n\alpha ^{2}ce^{n\alpha t}\left[ ce^{n\alpha t}-1\right]
^{\frac{2}{n}-2}>1~$and $\alpha ^{2}(2ce^{n\alpha t}-n)ce^{n\alpha t}\left[
ce^{n\alpha t}-1\right] ^{\frac{2}{n}-2}>1$. The following figures show the
dynamical behavior of energy densities of radiation and dark energy (i.e.
cosmological constant) in the early universe. With the same choice of the
model parameters ($c=1.2$, $n=1.23$, $1.43$, $1.63$, $\alpha =0.08$, $0.10$, 
$0.12$) with suitable time units, we have graphically represented the time
evolution of $\rho $ and $\rho _{de}$ for flat ($k=0$) and closed ($k=1$)
cases only. (The above numerical choice of $n$, $c$ and $\alpha $ are not
suitable for open case ($k=-1$)).\newline

\begin{figure}[tph]
\begin{center}
{\scriptsize $%
\begin{array}{c@{\hspace{.1in}}cc}
\includegraphics[width=2in]{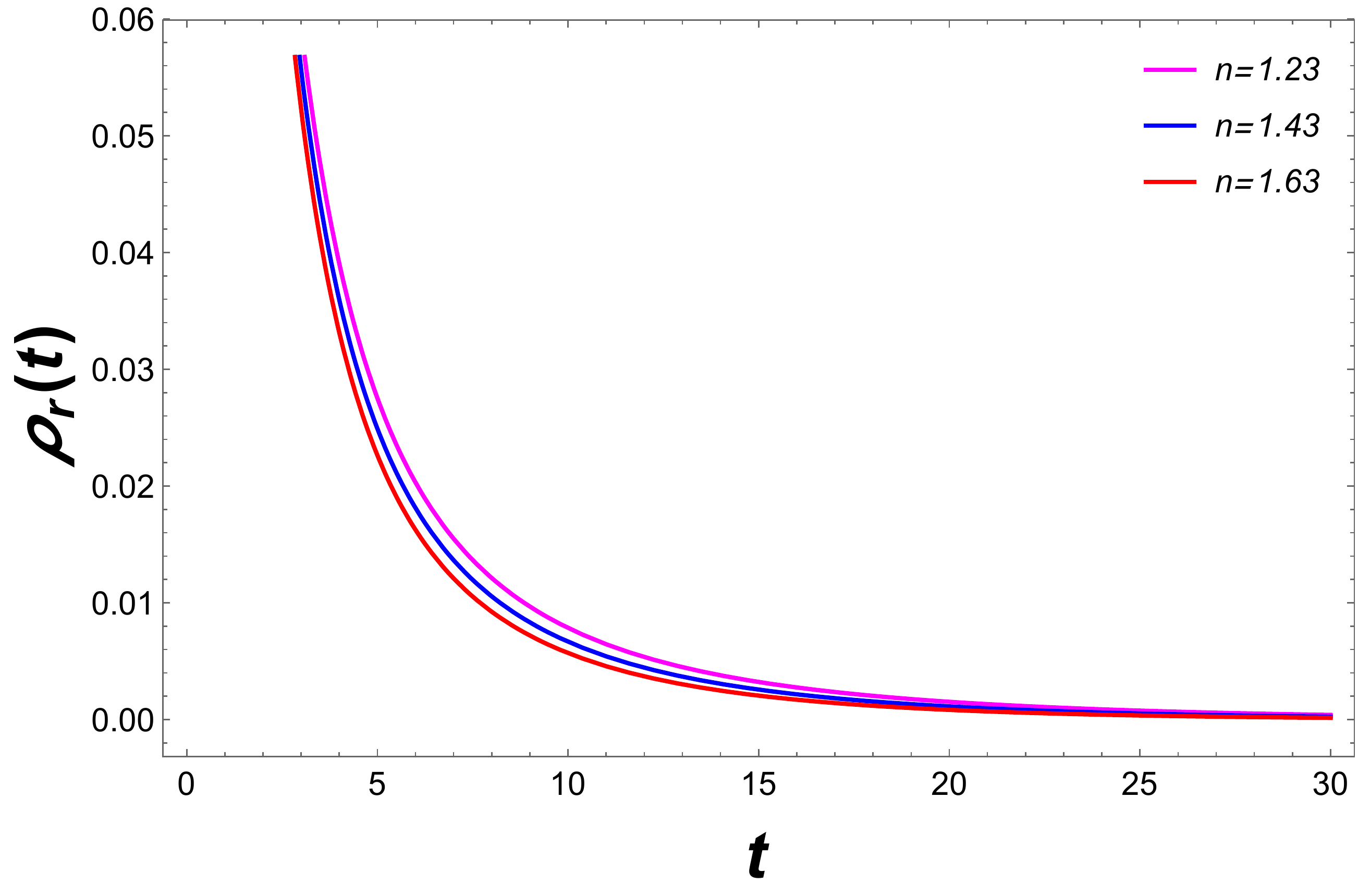} & %
\includegraphics[width=2in]{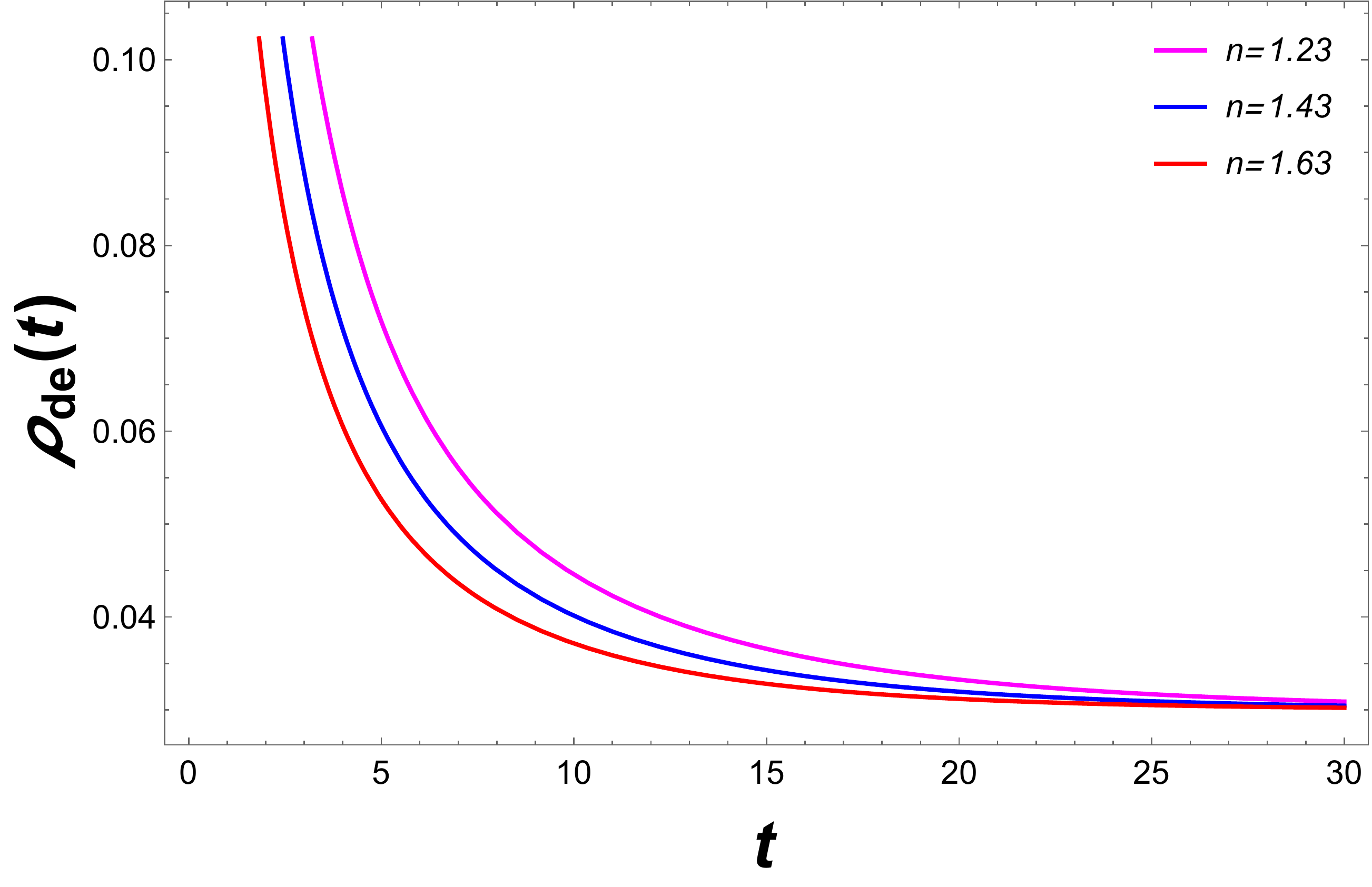} &  \\ 
\mbox (a) & \mbox (b) & 
\end{array}%
$ }
\end{center}
\caption{The plot shows a sketch of the time evolution of the physical
parameters (a) radiation energy density `$\protect\rho _{r}$' (b) dark
energy density `$\protect\rho _{de}$' for flat ($k=0$) case with suitable
units of cosmic time `$t$'. For these plots, we have chosen the integrating
constant $c=1.2$, the model parameter $\protect\alpha =0.1$ is fixed with
different values of $n=1.23,1.43,1.63$.}
\label{fig:3}
\end{figure}

\begin{figure}[tph]
\begin{center}
{\scriptsize $%
\begin{array}{c@{\hspace{.1in}}cc}
\includegraphics[width=2in]{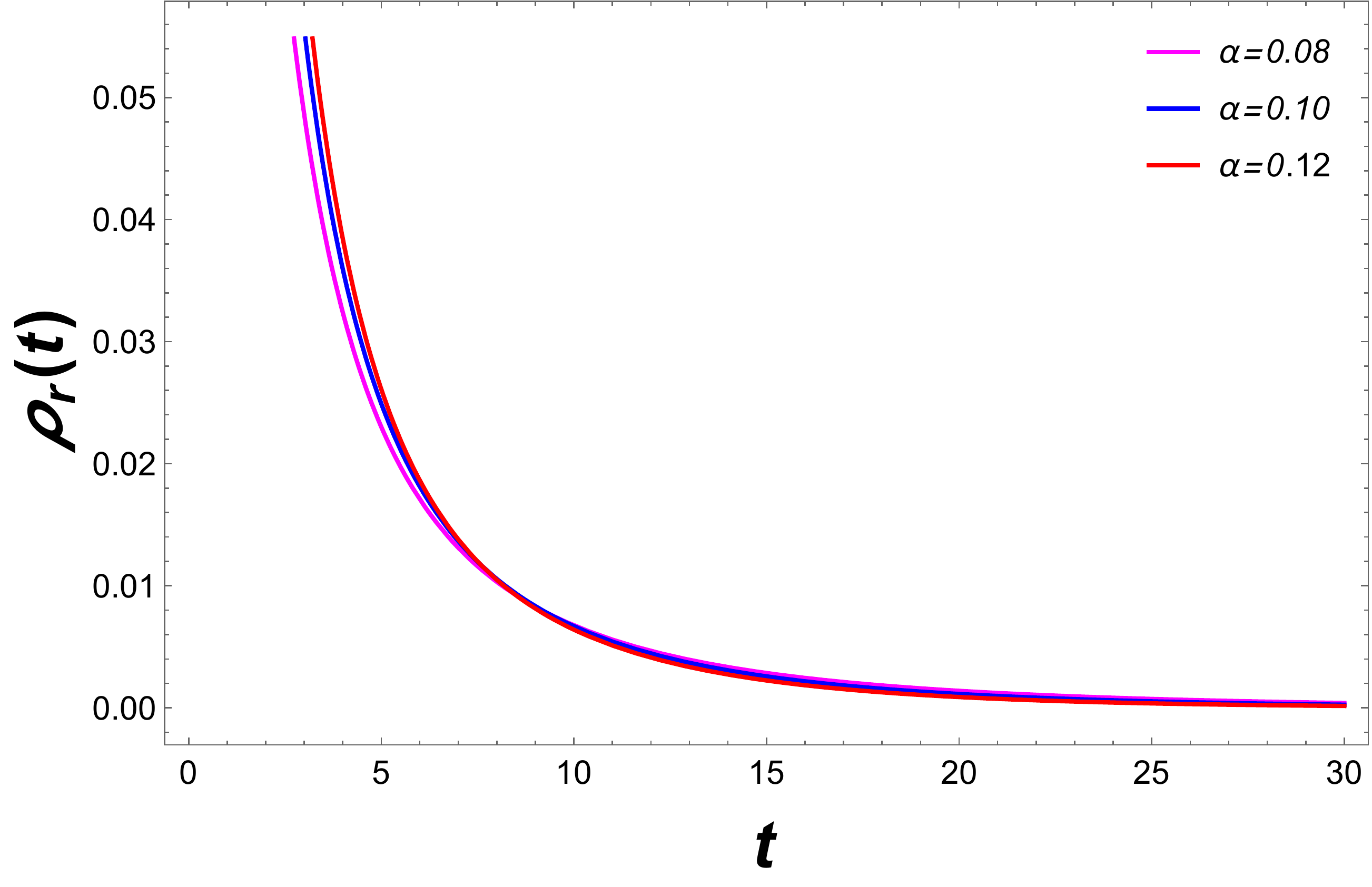} & %
\includegraphics[width=2in]{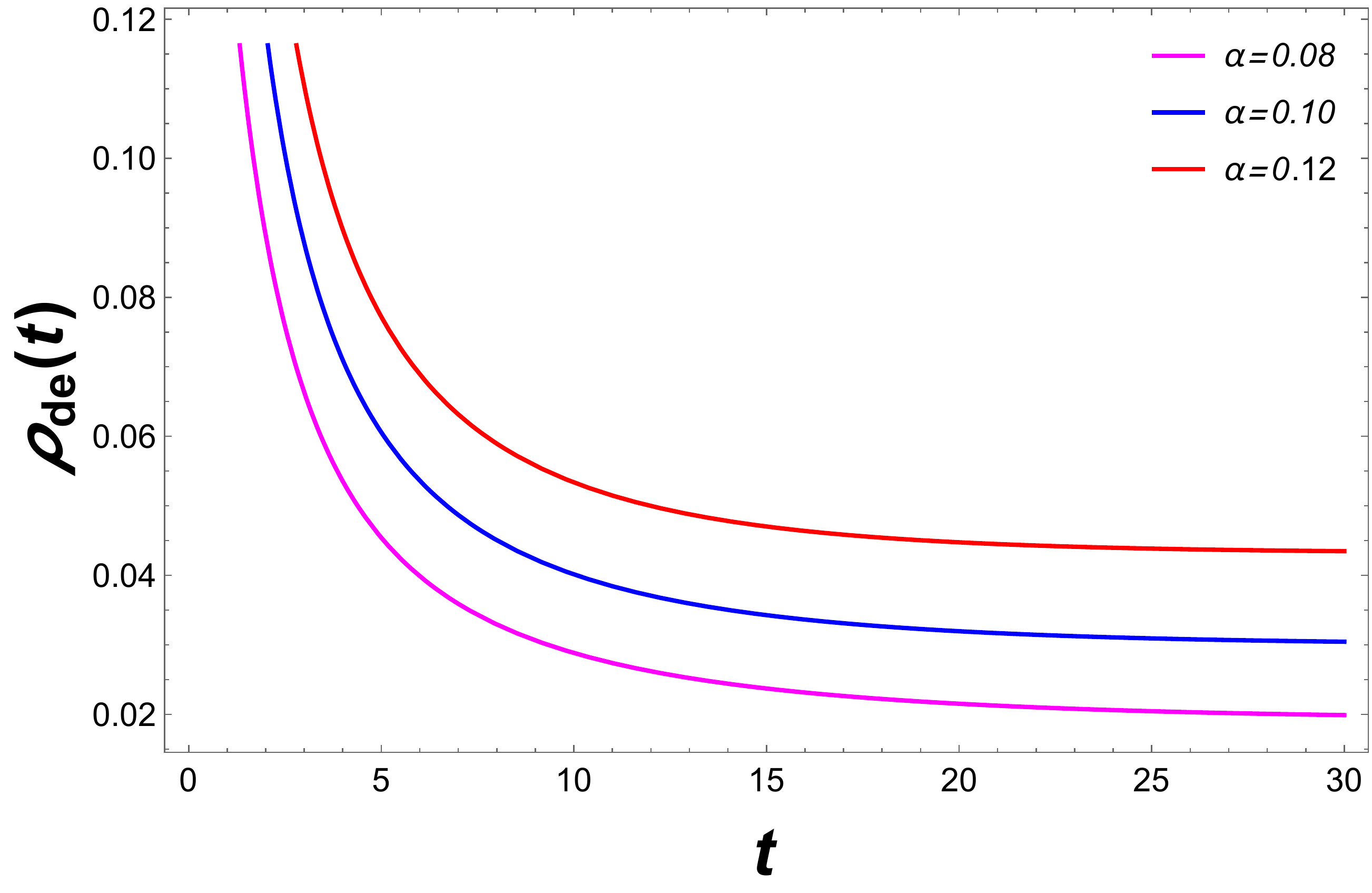} &  \\ 
\mbox (a) & \mbox (b) & 
\end{array}%
$ }
\end{center}
\caption{ The plot shows a sketch of the time evolution of the physical
parameters (a) radiation energy density `$\protect\rho _{r}$' (b) dark
energy density `$\protect\rho _{de}$' for flat ($k=0$) case with suitable
units of cosmic time `$t$'. For these plots, we have chosen the integrating
constant $c=1.2$, the model parameter $n=1.43$ is fixed with different
values of $\protect\alpha =0.08,0.10,0.12$.}
\label{fig:4}
\end{figure}

\begin{figure}[tph]
\begin{center}
{\scriptsize $%
\begin{array}{c@{\hspace{.1in}}cc}
\includegraphics[width=2in]{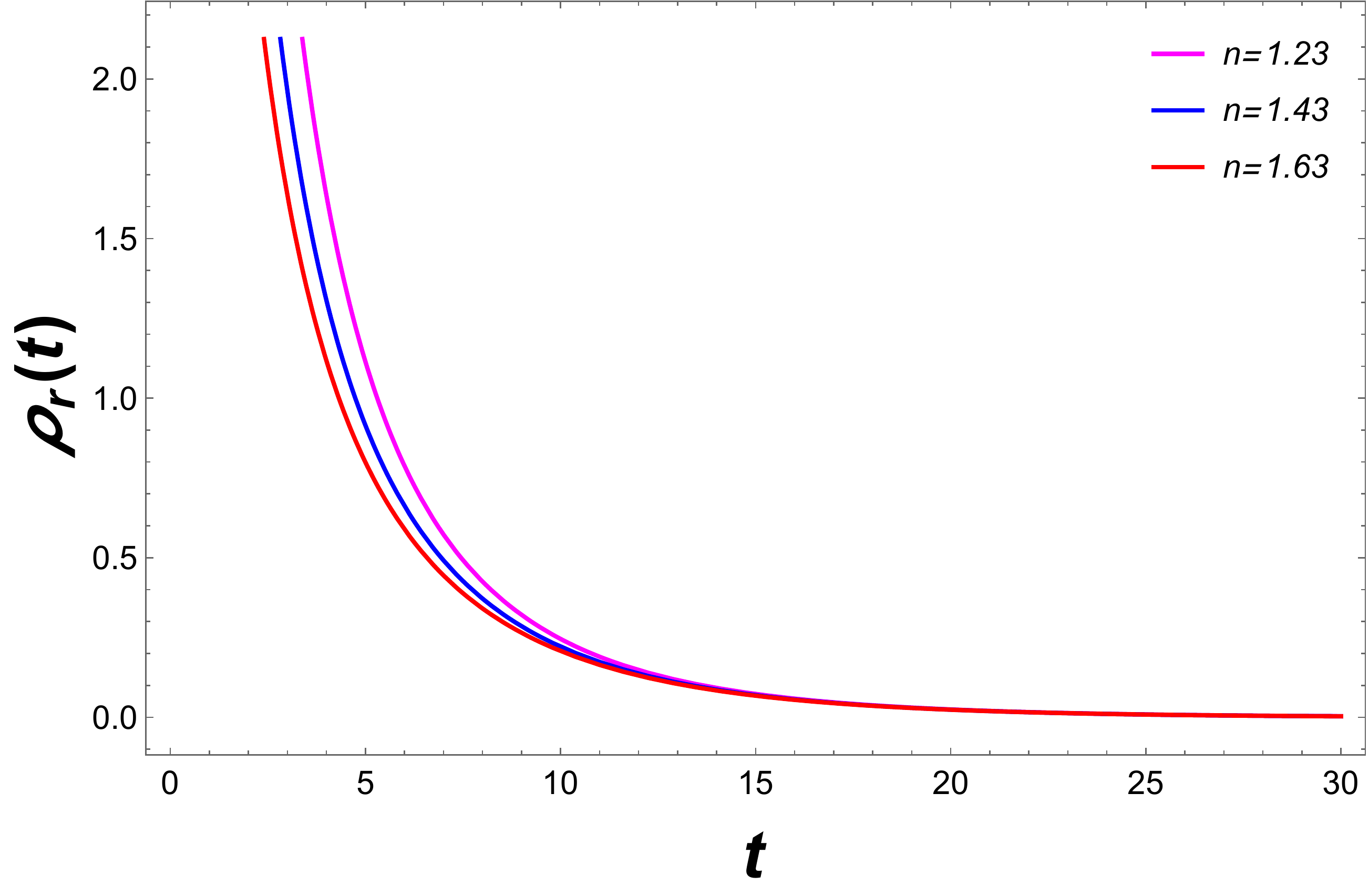} & %
\includegraphics[width=2in]{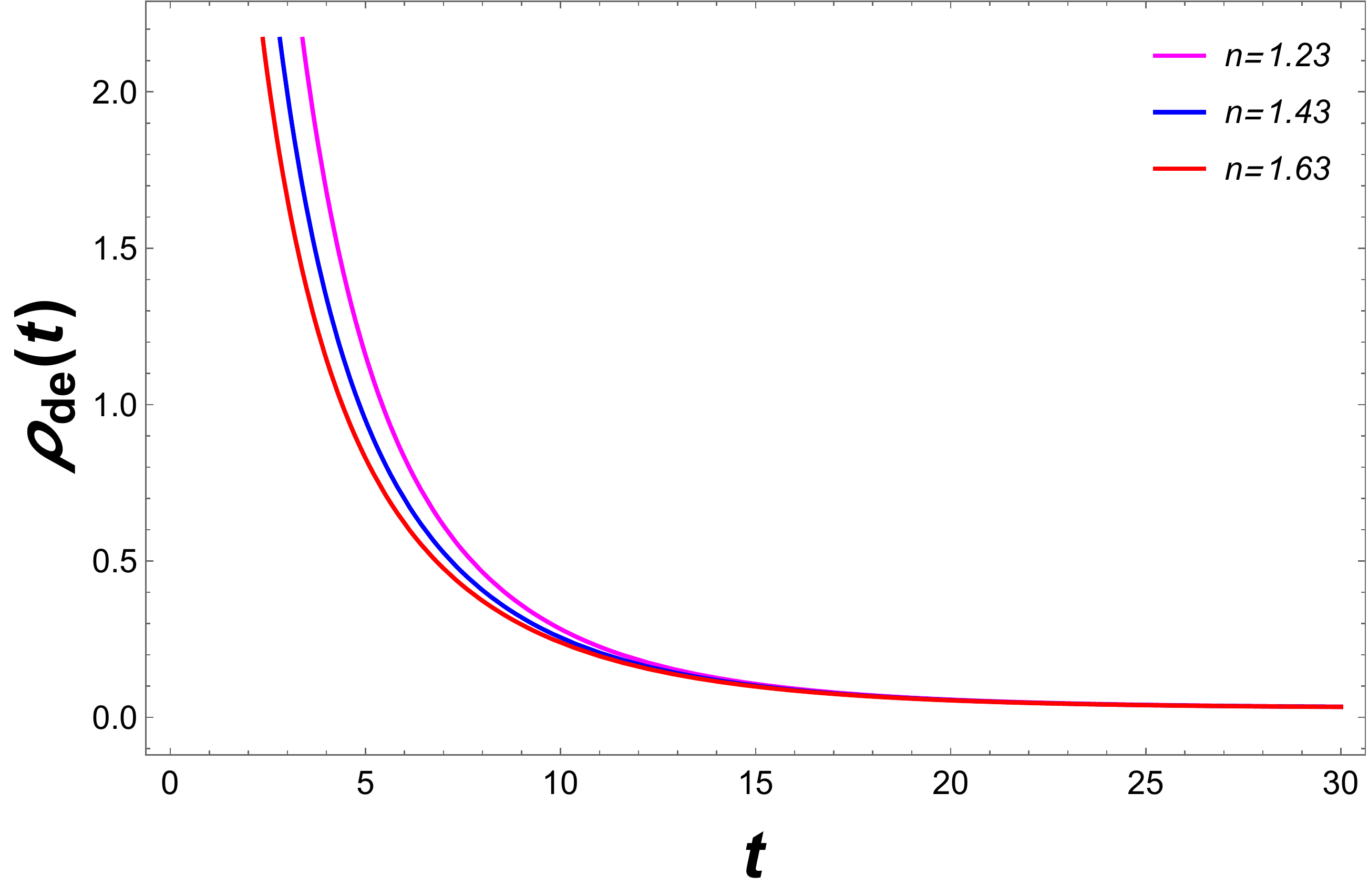} &  \\ 
\mbox (a) & \mbox (b) & 
\end{array}%
$ }
\end{center}
\caption{ The plot shows a sketch of the time evolution of the physical
parameters (a) radiation energy density `$\protect\rho _{r}$' (b) dark
energy density `$\protect\rho _{de}$' for closed ($k=1$) case with suitable
units of cosmic time `$t$'. For these plots, we have chosen the integrating
constant $c=1.2$, the model parameter $\protect\alpha =0.1$ is fixed with
different values of $n=1.23,1.43,1.63$.}
\label{fig:5}
\end{figure}

\begin{figure}[tph]
\begin{center}
{\scriptsize $%
\begin{array}{c@{\hspace{.1in}}cc}
\includegraphics[width=2in]{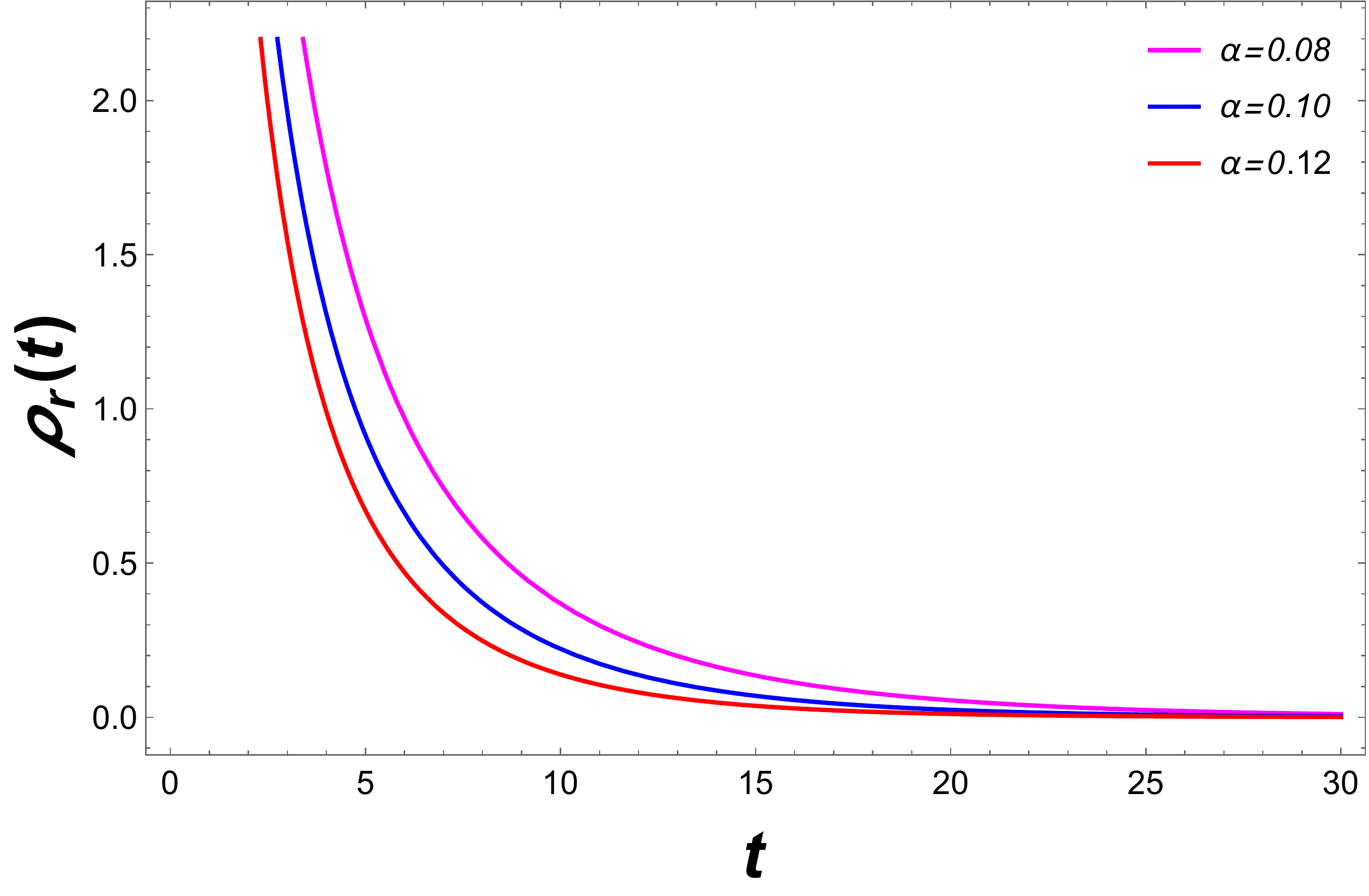} & %
\includegraphics[width=2in]{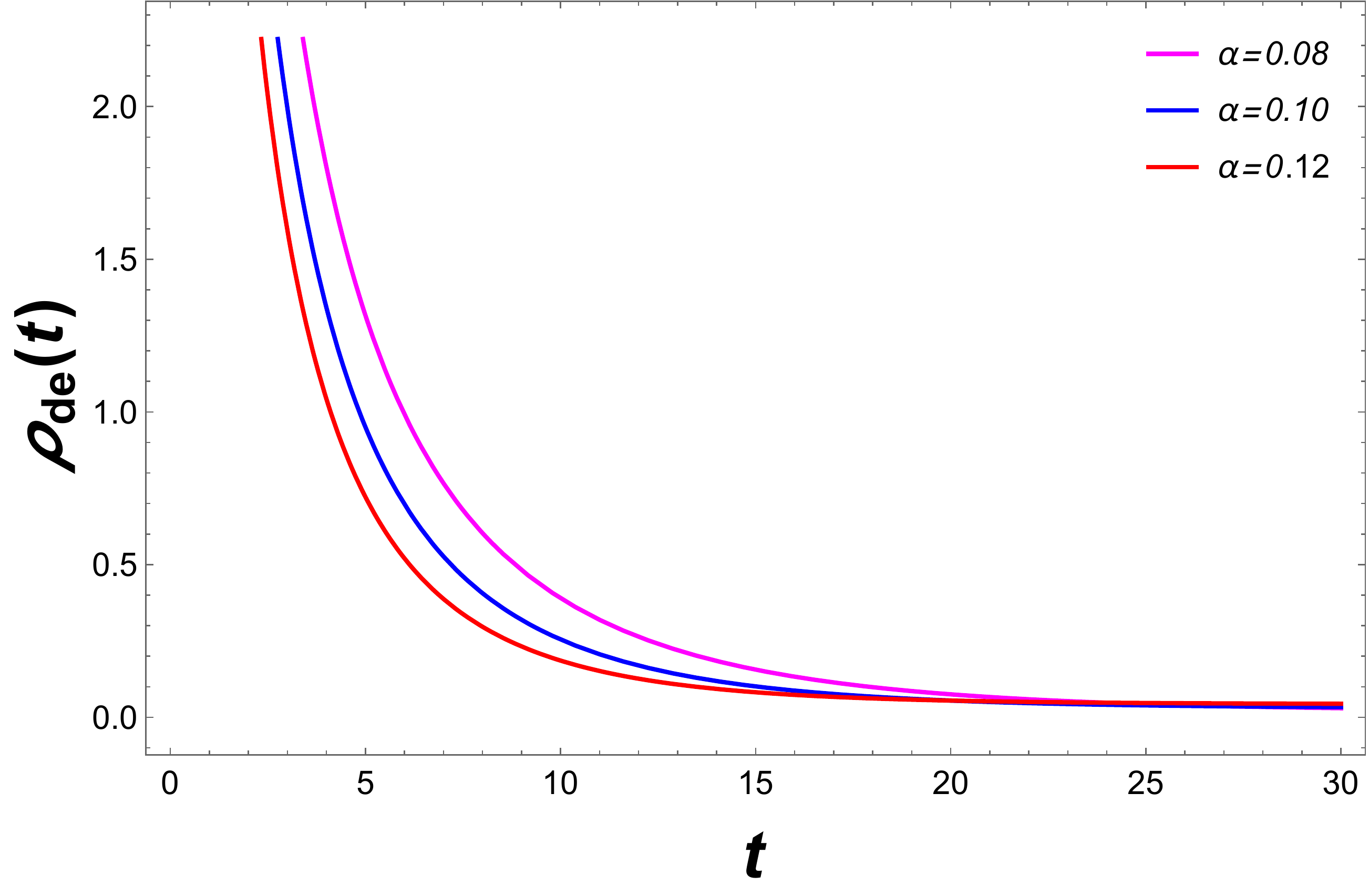} &  \\ 
\mbox (a) & \mbox (b) & 
\end{array}%
$ }
\end{center}
\caption{ The plot shows a sketch of the time evolution of the physical
parameters (a) radiation energy density `$\protect\rho _{r}$' (b) dark
energy density `$\protect\rho _{de}$' for closed ($k=1$) case with suitable
units of cosmic time `$t$'. For these plots, we have chosen the integrating
constant $c=1.2$, the model parameter $n=1.43$ is fixed with different
values of $\protect\alpha =0.08,0.10,0.12$.}
\label{fig:6}
\end{figure}

\qquad We can observe that in all the cases, the energy densities of both
the radiation and dark energy are very high initially and decreases rapidly
as time unfolds by creating photons in the early universe. We know the
radiation energy density and the temperature ($T$) are related by the
relation 
\begin{equation}
\rho _{r}=\frac{\pi ^{2}}{30}N(T)T^{4},  \label{new8}
\end{equation}%
in the units with $k_{B}=c=$%
%TCIMACRO{\U{127}}%
%BeginExpansion
h{\hskip-.2em}\llap{\protect\rule[1.1ex]{.325em}{.1ex}}{\hskip.2em}%
%EndExpansion
$=1$. At temperature $T$, the effective number of spin degrees of freedom $%
N(T)$ is given by $N(T)=\frac{7}{8}N_{f}(T)+N_{b}(T)$, where $N_{f}(T)$ and $%
N_{b}(T)$ correspond to fermions and bosons respectively. We assume $N(T)$
to be constant throughout this era. From equations (\ref{new6}) and (\ref%
{new8}), we obtain 
\begin{equation}
T=\left( \frac{45}{\pi ^{2}N}\right) ^{\frac{1}{4}}\left[ \frac{n\alpha
^{2}ce^{n\alpha t}}{\left[ ce^{n\alpha t}-1\right] ^{2}}+\frac{k}{\left[
ce^{n\alpha t}-1\right] ^{\frac{2}{n}}}\right] ^{\frac{1}{4}},  \label{new9}
\end{equation}%
showing that in the beginning as $t\rightarrow 0$, we have $T^{(i)}\approx
\left( \frac{45}{\pi ^{2}N}\right) ^{\frac{1}{4}}\left[ n\alpha
^{2}c(c-1)^{-2}+k(c-1)^{-\frac{2}{n}}\right] ^{\frac{1}{4}}$ implying that
the radiation temperature also attains a finite value initially. The
following figures show the variation of radiation temperature in the early
universe with the same choice of model parameters. (Here also, these
numerical choice of model parameters are not suitable for open $k=-1$ case.)

\begin{figure}[tph]
\begin{center}
{\scriptsize $%
\begin{array}{c@{\hspace{.1in}}cc}
\includegraphics[width=2in]{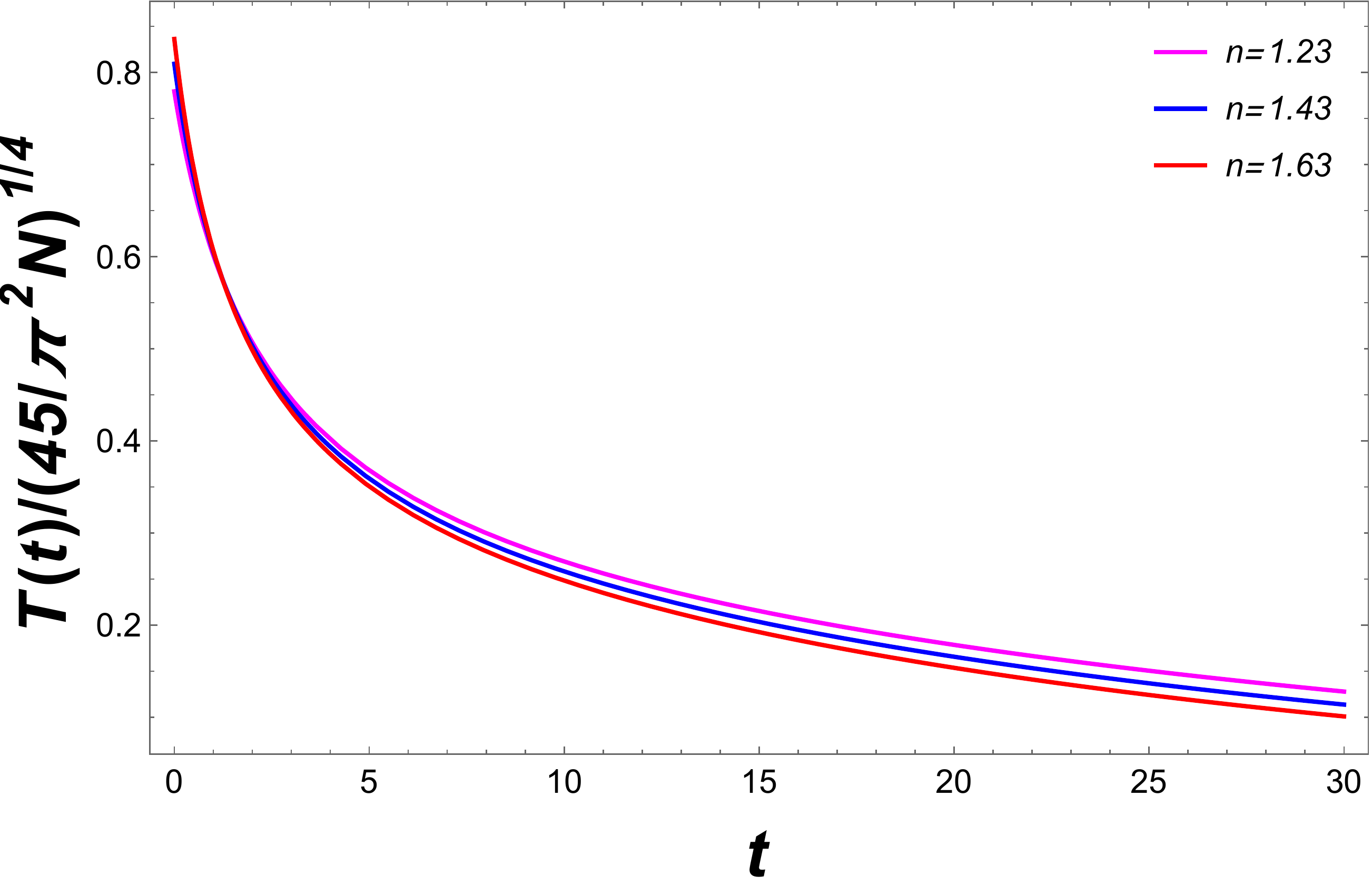} & %
\includegraphics[width=2in]{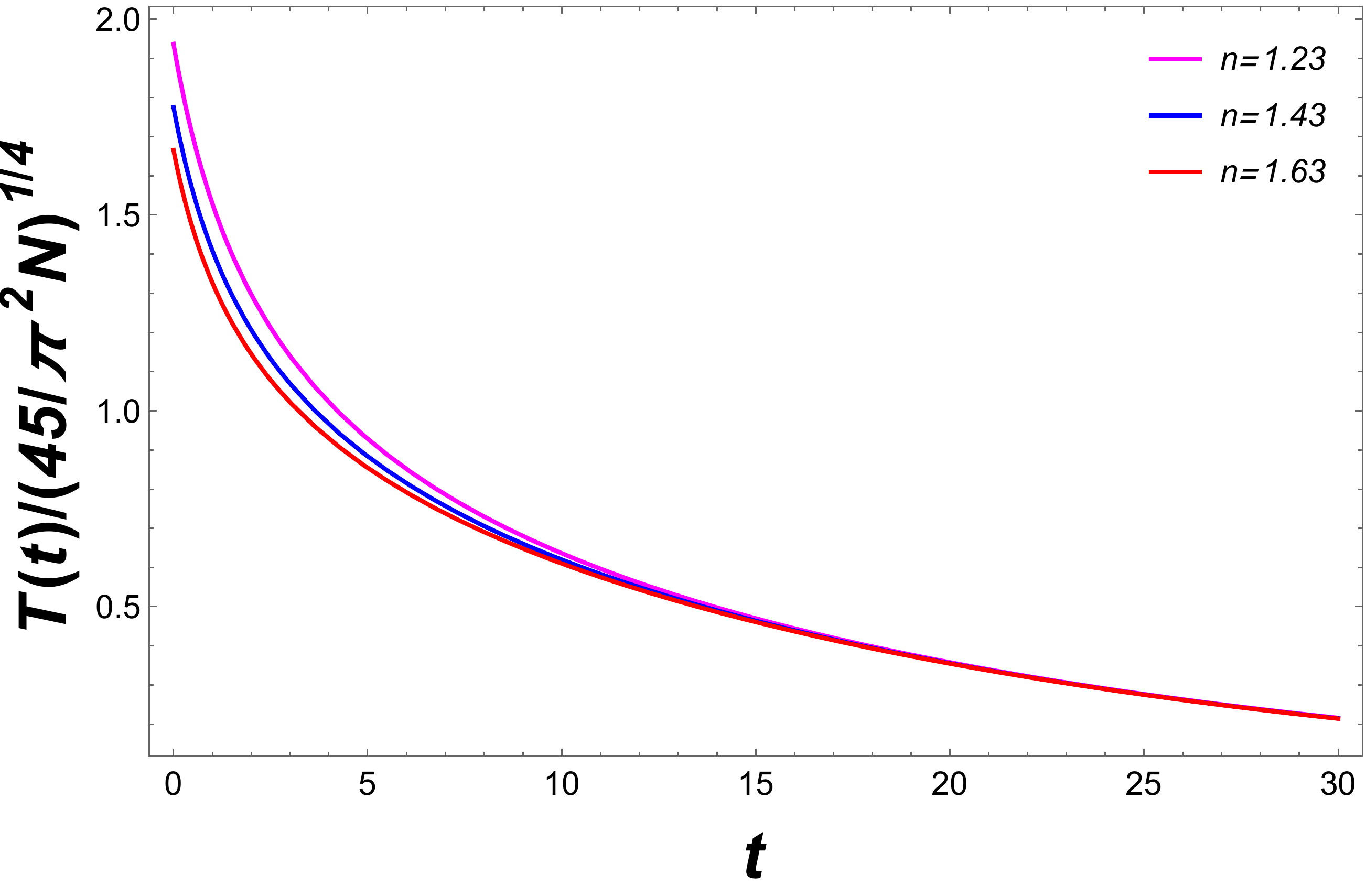} &  \\ 
\mbox (a) & \mbox (b) & 
\end{array}%
$ }
\end{center}
\caption{ The plot shows a sketch of the time evolution of the radiation
temperature `$T$' for (a) flat case ($k=0$) (b) closed case ($k=1$) with
suitable units of cosmic time `$t$'. For these plots, we have chosen the
integrating constant $c=1.2$, the model parameter $\protect\alpha =0.1$ is
fixed with different values of $n=1.23,1.43,1.63$.}
\label{fig:7}
\end{figure}

\begin{figure}[tph]
\begin{center}
{\scriptsize $%
\begin{array}{c@{\hspace{.1in}}cc}
\includegraphics[width=2in]{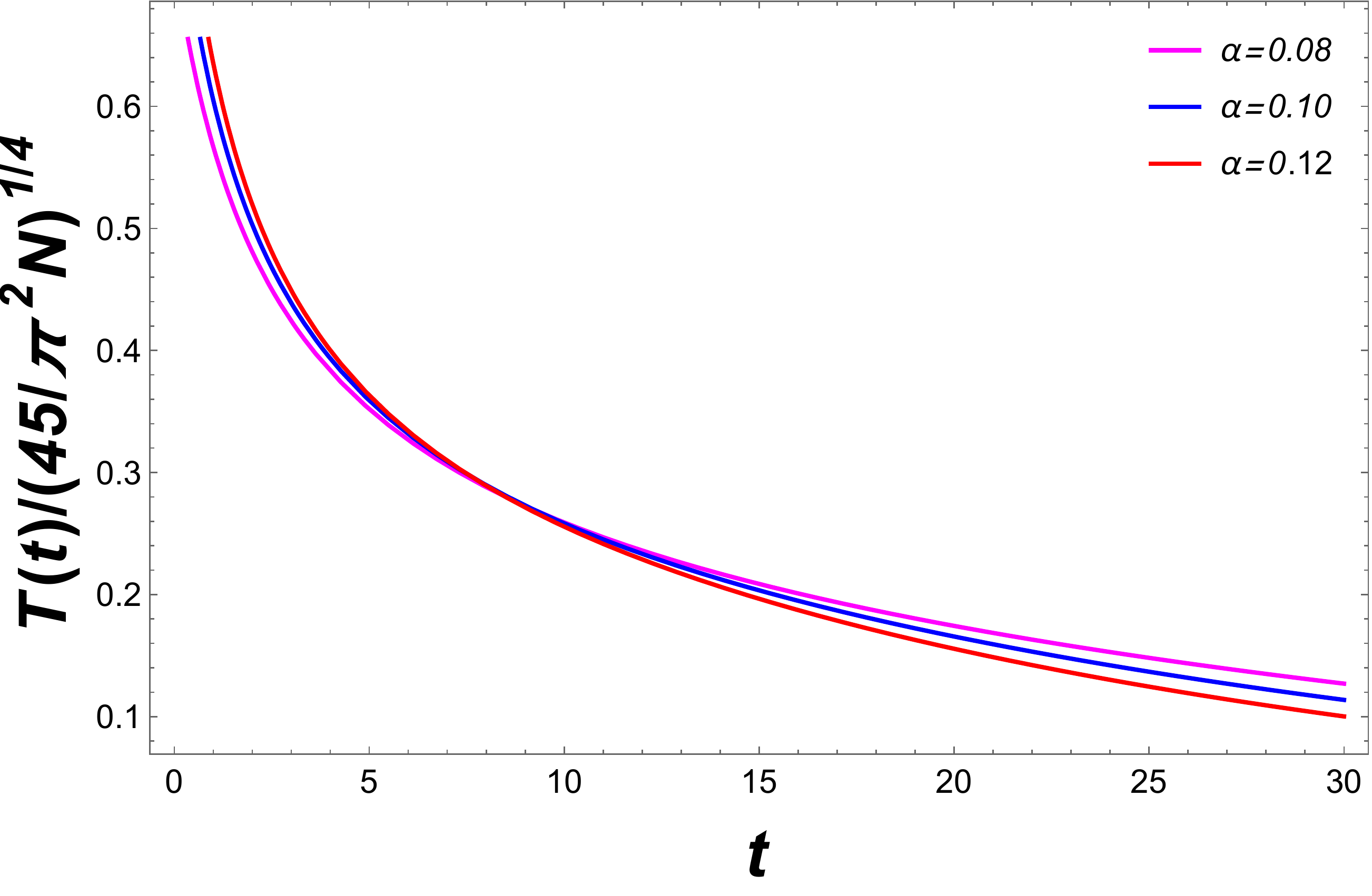} & %
\includegraphics[width=2in]{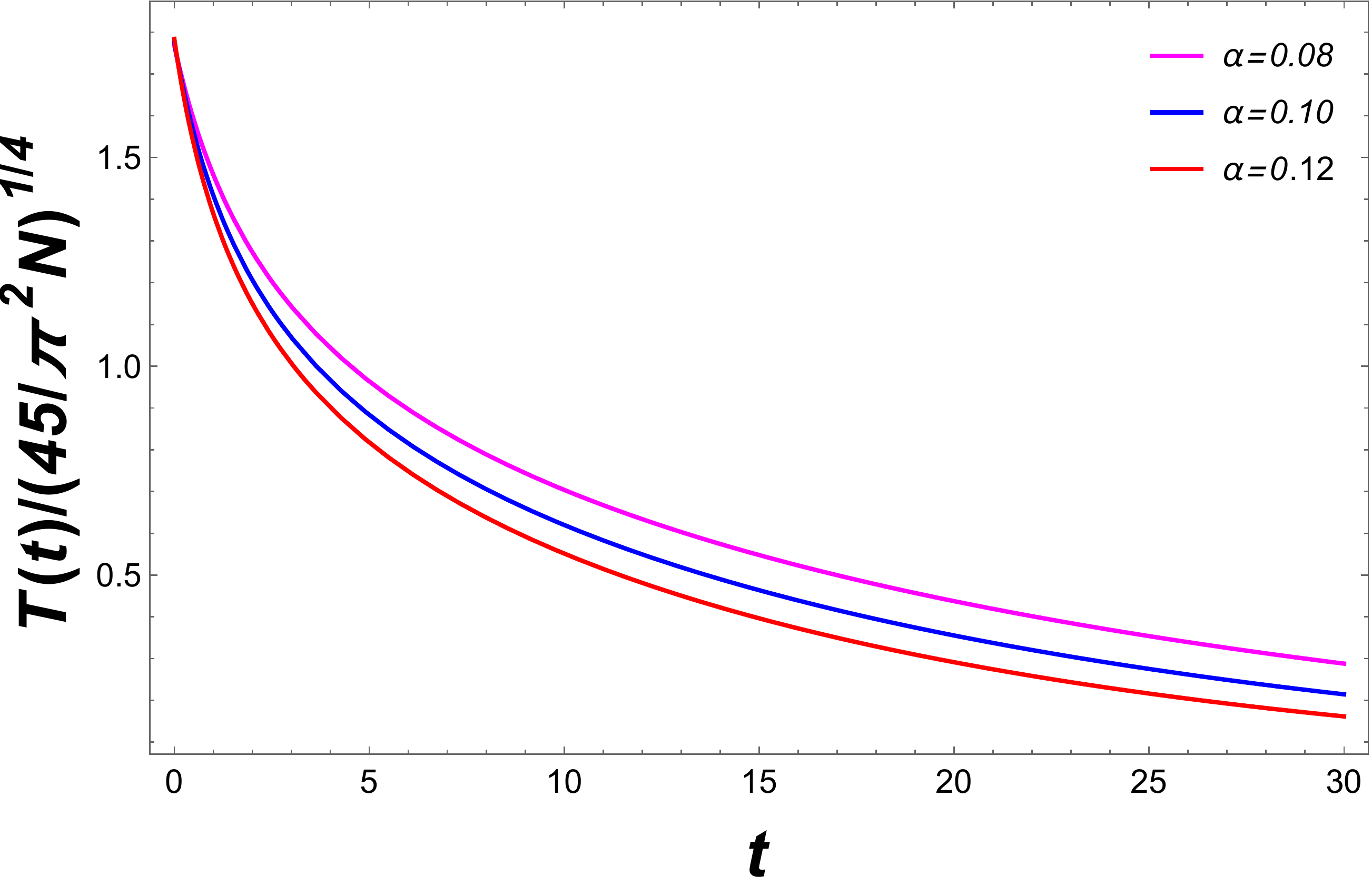} &  \\ 
\mbox (a) & \mbox (b) & 
\end{array}%
$ }
\end{center}
\caption{ The plot shows a sketch of the time evolution of the radiation
temperature `$T$' for (a) flat case ($k=0$) (b) closed case ($k=1$) with
suitable units of cosmic time `$t$'. For these plots, we have chosen the
integrating constant $c=1.2$, the model parameter $n=1.43$ is fixed with
different values of $\protect\alpha =0.08,0.10,0.12$.}
\label{fig:8}
\end{figure}

\subsection{Matter Dominated universe}

\qquad In the late \textit{matter dominated era}, we have $w=0$ and $\rho
\approx \rho _{m}$. Equations (\ref{new4}) and (\ref{new5}) together with (%
\ref{sf1}) reduce to 
\begin{eqnarray}
\rho _{m} &=&2\left[ \frac{n\alpha ^{2}ce^{n\alpha t}}{\left[ ce^{n\alpha
t}-1\right] ^{2}}+\frac{k}{\left[ ce^{n\alpha t}-1\right] ^{\frac{2}{n}}}%
\right] ,  \label{new10} \\
\rho _{de} &=&\left[ \frac{\alpha ^{2}\left( 3ce^{n\alpha t}-2n\right)
ce^{n\alpha t}}{\left[ ce^{n\alpha t}-1\right] ^{2}}+\frac{k}{\left[
ce^{n\alpha t}-1\right] ^{\frac{2}{n}}}\right] .  \label{new11}
\end{eqnarray}

As $t\rightarrow \infty $, $\rho _{m}\rightarrow 0$ and $\rho
_{de}\rightarrow 3\alpha ^{2}$ which is constant. In order to study the late
time behavior of these cosmological parameters, it will be better to express
them in terms of redshift ($1+z=\frac{a_{0}}{a}$). For simplicity we
normalize the case and consider the present value of scale factor to be $1$
(i.e. $a_{0}=1$). To do that, we establish the $t-z$ relationship here which
comes out to be $t(z)=\frac{1}{n\alpha }\log \left[ c^{-1}\{1+(1+z)^{-n}\}%
\right] $. So, the Hubble parameter $H$ can be written in terms of redshift
as 
\begin{equation}
H(z)=\alpha \left[ 1+(1+z)^{n}\right] ,  \label{11a}
\end{equation}%
or, 
\begin{equation}
H(z)=\frac{H_{0}}{2}\left[ 1+(1+z)^{n}\right] ,  \label{11b}
\end{equation}%
where $H_{0}$ is the present value of the Hubble parameter. The following
figures FIG.\ref{fig:9} \& FIG.\ref{fig:10} show the dynamical behavior of
the energy densities in near past and late-time universe. The plots are in
terms of redshift $z$. In all cases they are decreasing to very small values.

\begin{figure}[tph]
\begin{center}
{\scriptsize $%
\begin{array}{c@{\hspace{.1in}}cc}
\includegraphics[width=2in]{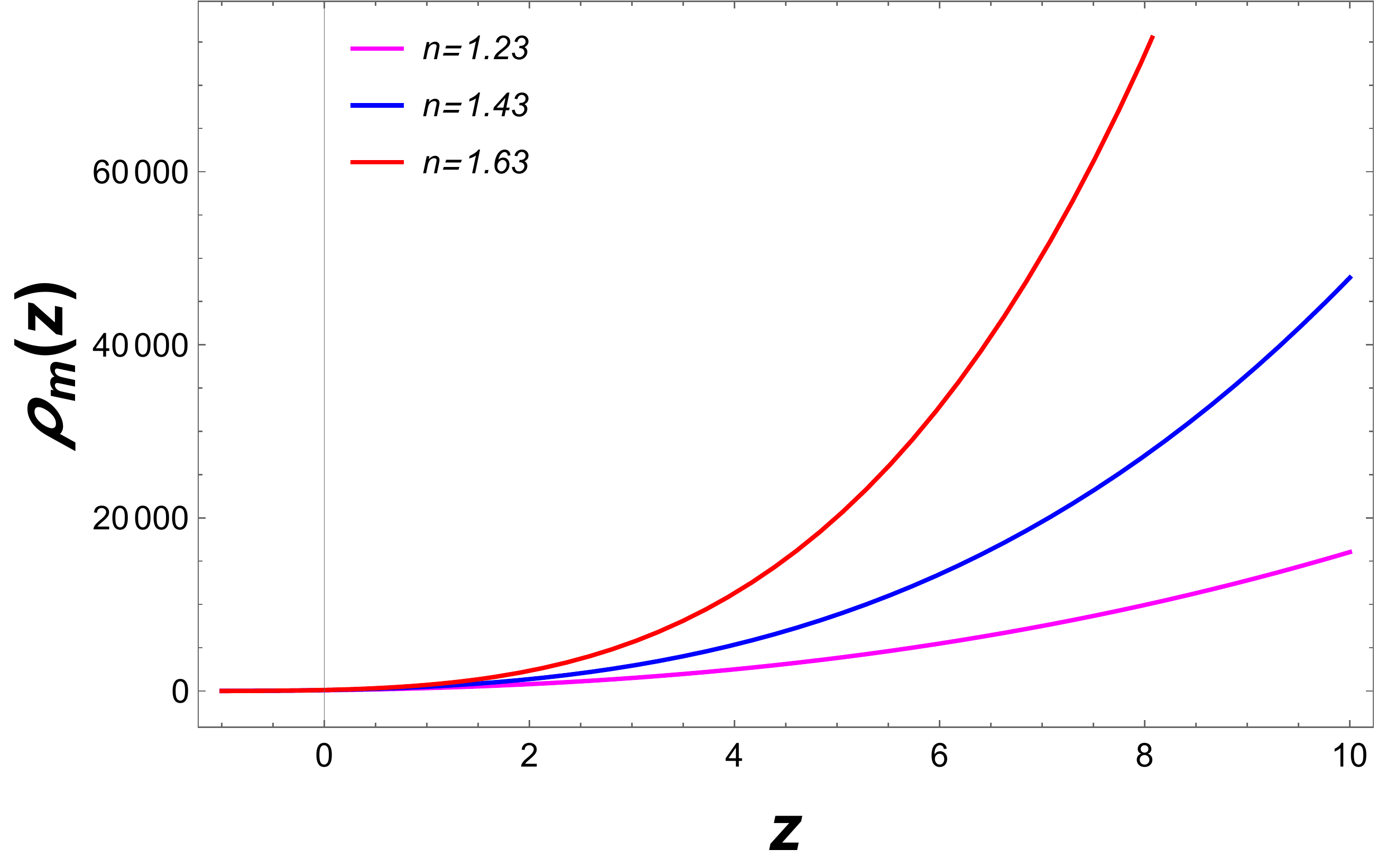} & %
\includegraphics[width=2in]{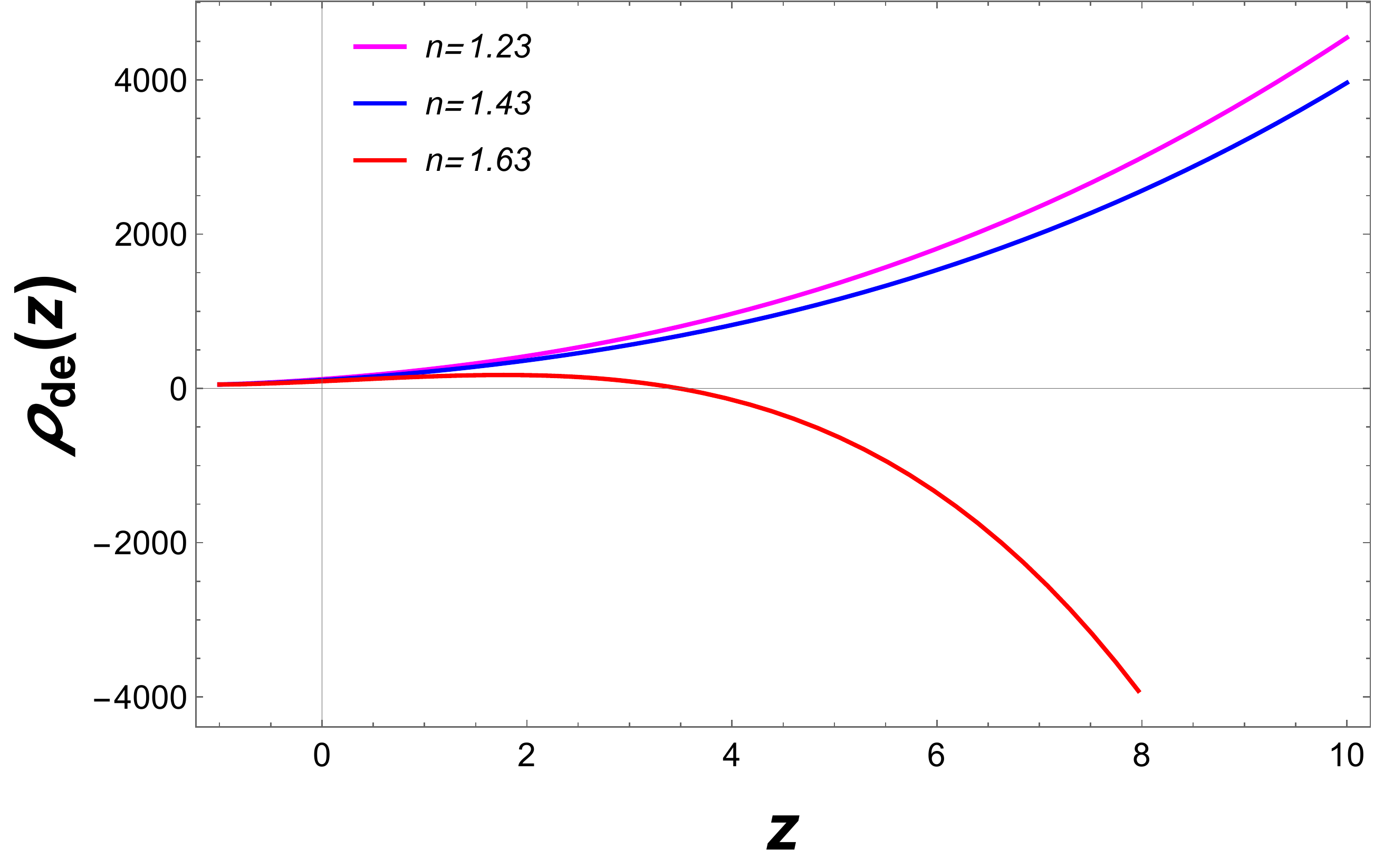} &  \\ 
\mbox (a) & \mbox (b) & 
\end{array}%
$ }
\end{center}
\caption{ The plot shows a sketch of the evolution of (a) matter energy
density `$\protect\rho _{m}(z)$' (b) dark energy density `$\protect\rho %
_{de}(z)$' for flat ($k=0$) case w.r.t resdshift `$z$'. The only model
parameter involved is $n$ and is chosen $n=1.43$.}
\label{fig:9}
\end{figure}

\begin{figure}[tph]
\begin{center}
{\scriptsize $%
\begin{array}{c@{\hspace{.1in}}cc}
\includegraphics[width=2in]{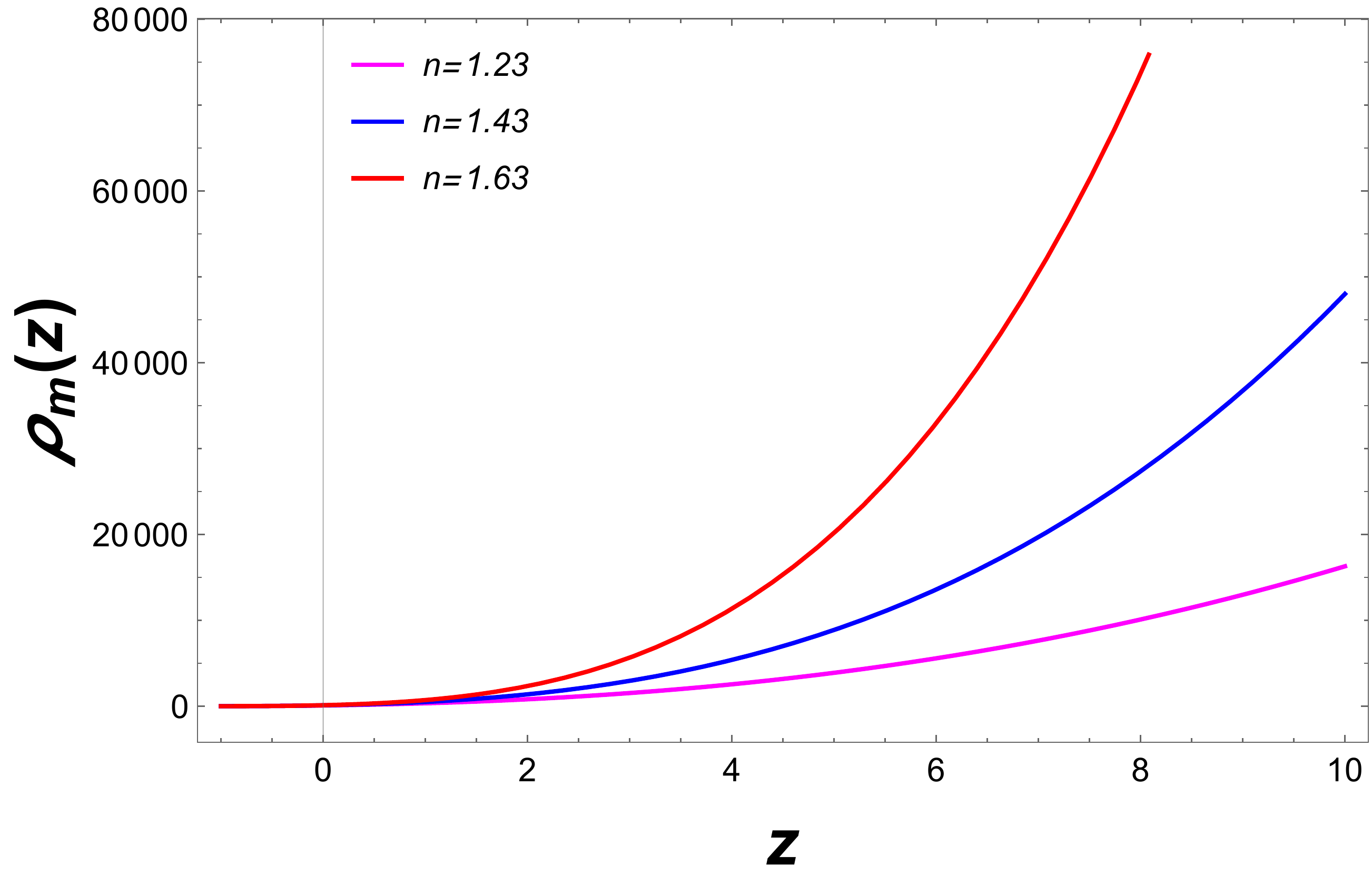} & %
\includegraphics[width=2in]{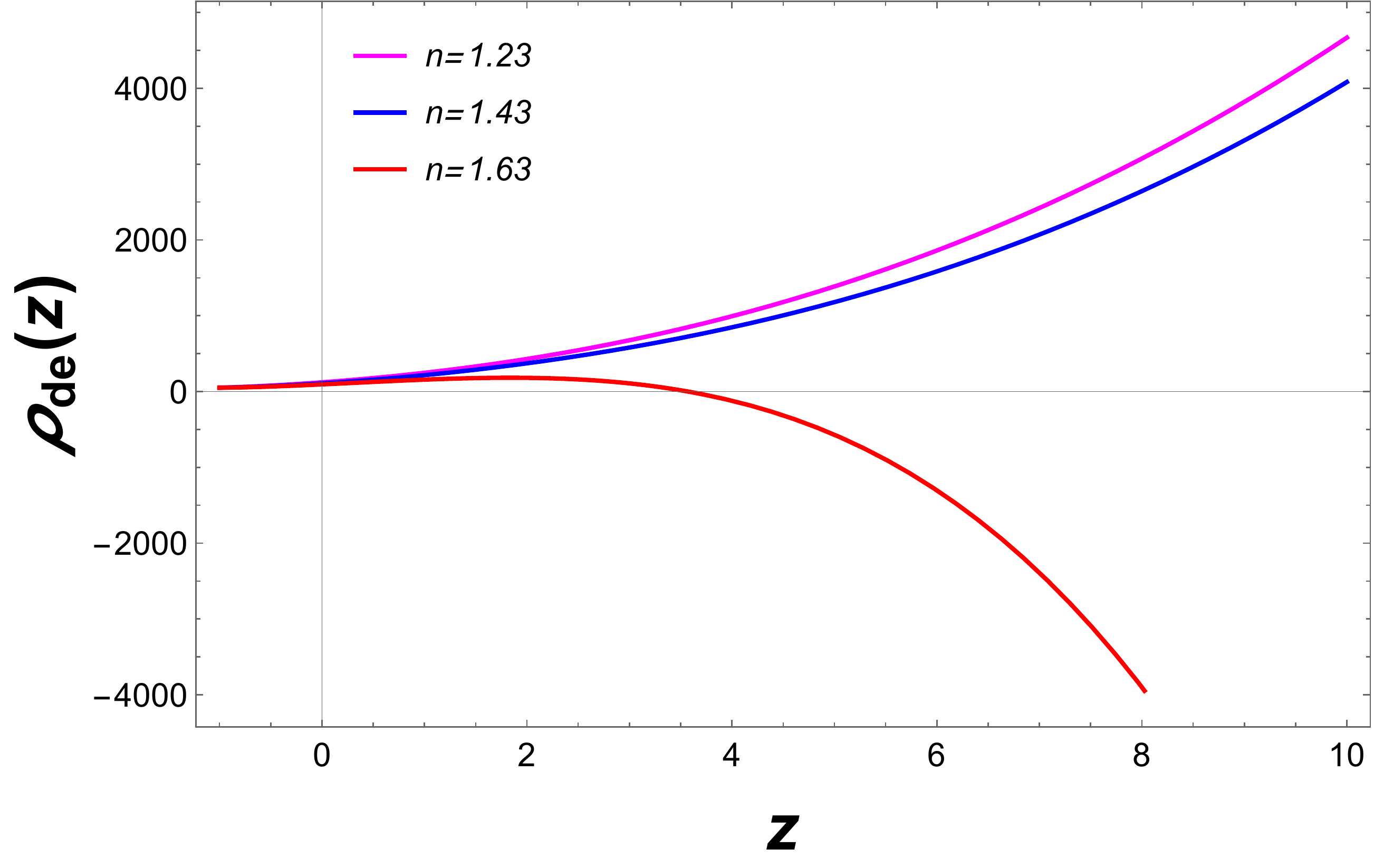} &  \\ 
\mbox (a) & \mbox (b) & 
\end{array}%
$ }
\end{center}
\caption{ The plot shows a sketch of the evolution of (a) matter energy
density `$\protect\rho _{m}(z)$' (b) dark energy density `$\protect\rho %
_{de}(z)$' for closed ($k=1$) case w.r.t resdshift `$z$'. The only model
parameter involved is $n$ and is chosen $n=1.43$.}
\label{fig:10}
\end{figure}
From these figures, we observe that the energy density of dark energy is
negative in the past for $n=1.63$ for both flat and closed cases but remains
positive for $n=1.23$, $1.43$ implying that the value of the model parameter
must be chosen carefully for which we constrain the value of $n$ with any
observational datasets. In the following section, we discuss the phase
transition scenario and perform the observational analysis.

\subsection{Deceleration to acceleration phase transition \& $H(z)$
Observation}

\qquad The parametrization of Hubble parameter we considered here, shows a
signature flip from early decelerating phase to late accelerating phase.
Recent observation depict that the phase transition occurred around $%
z\approx 0.7$. The choice of the model parameter $n=1.43$ is in good
agreement with this. We can plot the deceleration parameter w.r.t. redshift $%
z$ to have a better understanding on the phase transition of the obtained
model. The following figure shows that the obtained model had undergone from
an early decelerating phase to a late-time accelerating phase. We can also
plot the normalized Hubble parameter $E(z)$ ($=H/H_{0}$).

\begin{figure}[tph]
\begin{center}
{\scriptsize $%
\begin{array}{c@{\hspace{.1in}}cc}
\includegraphics[width=2in]{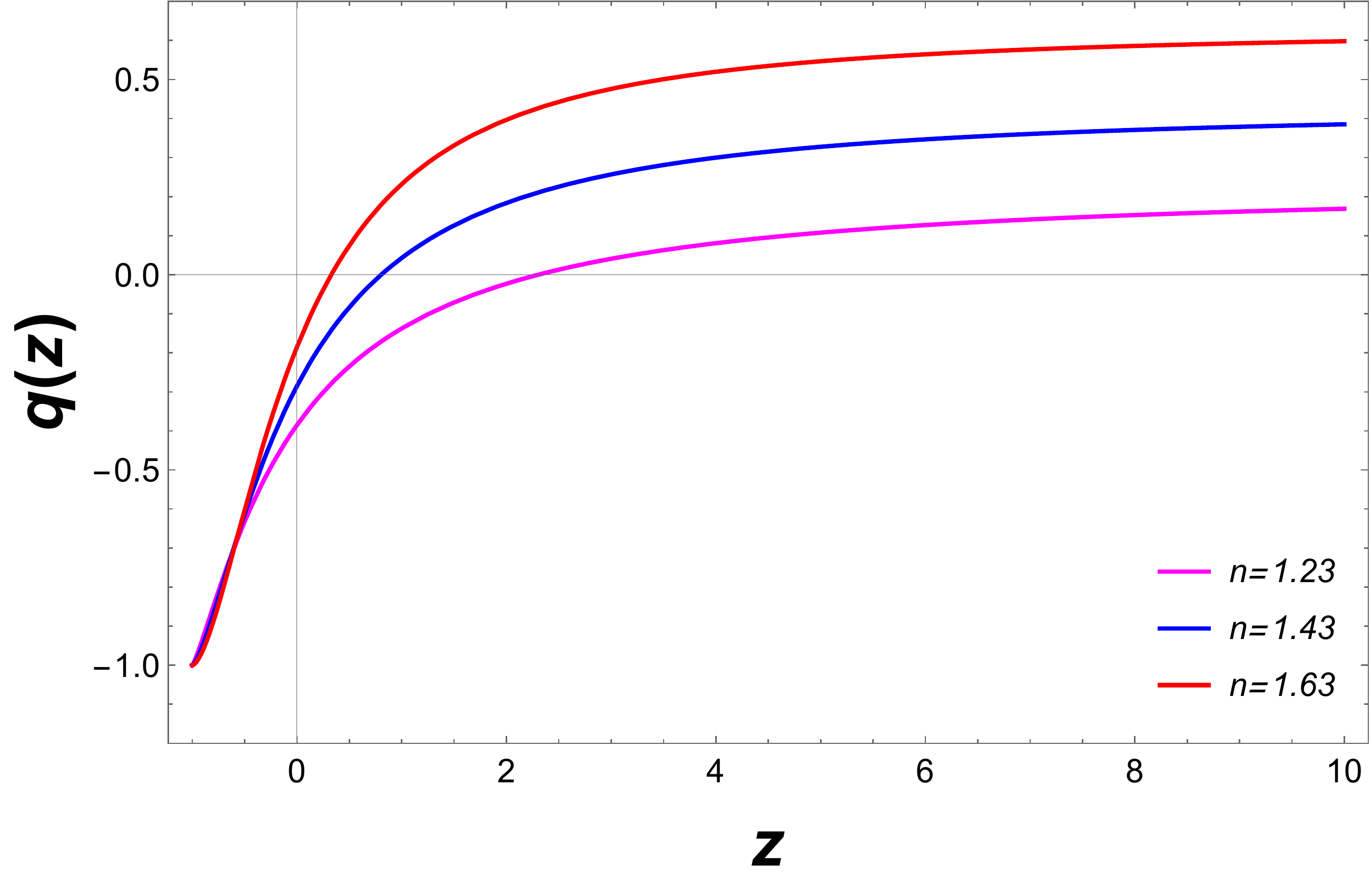} & \includegraphics[width=2in]{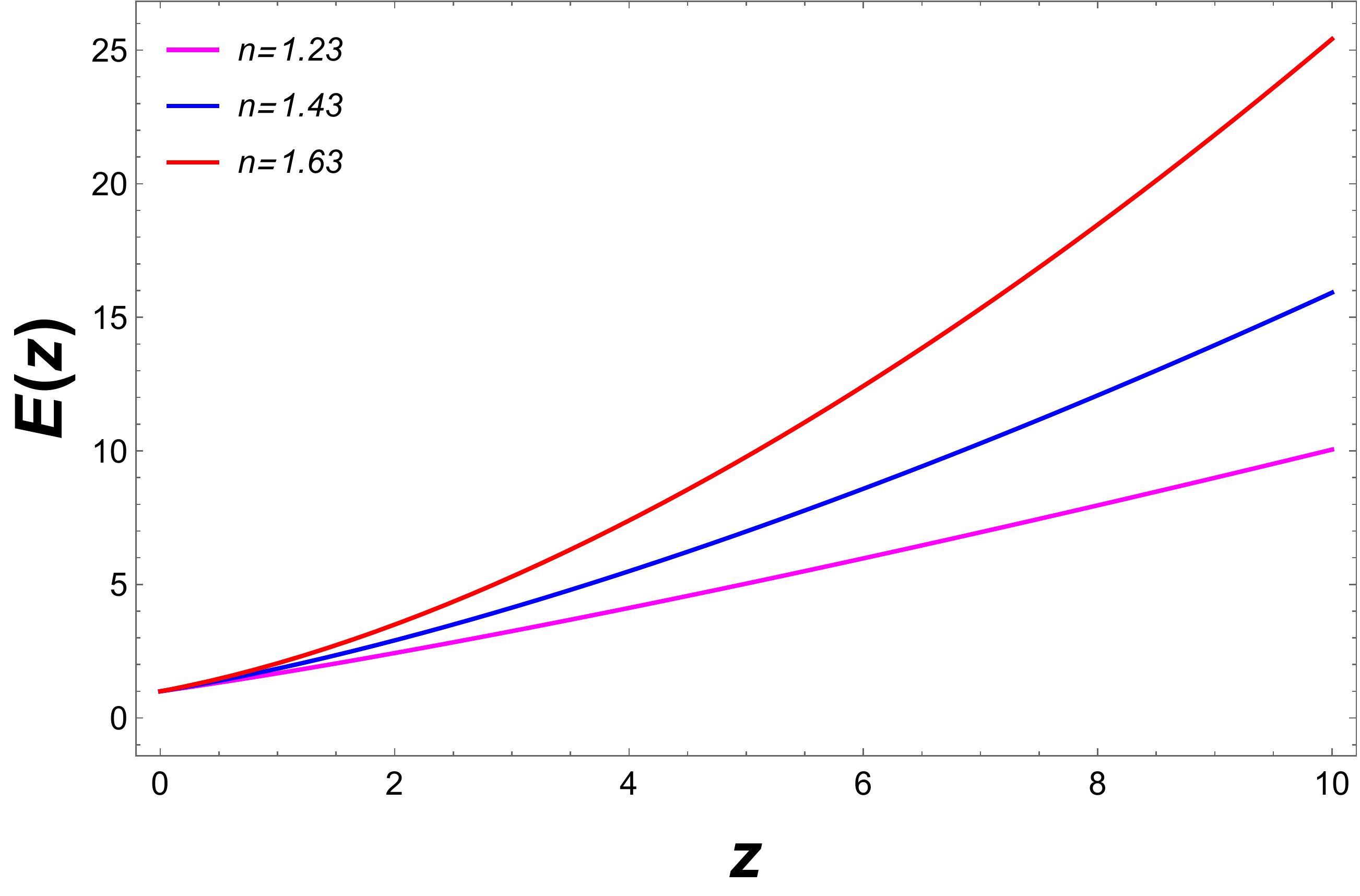} & 
\\ 
\mbox (a) & \mbox (b) & 
\end{array}%
$ }
\end{center}
\caption{ The plot shows a sketch of the evolution of (a) deceleration
parameter `$q(z)$' (b) normalized Hubble parameter `$E(z)$' w.r.t resdshift `%
$z$'. The only model parameter involved in $n$ and is chosen $n=1.43$.}
\label{fig:11}
\end{figure}

\begin{figure}[tph]
\begin{center}
{\scriptsize $%
\begin{array}{c@{\hspace{.1in}}cc}
\includegraphics[width=2in]{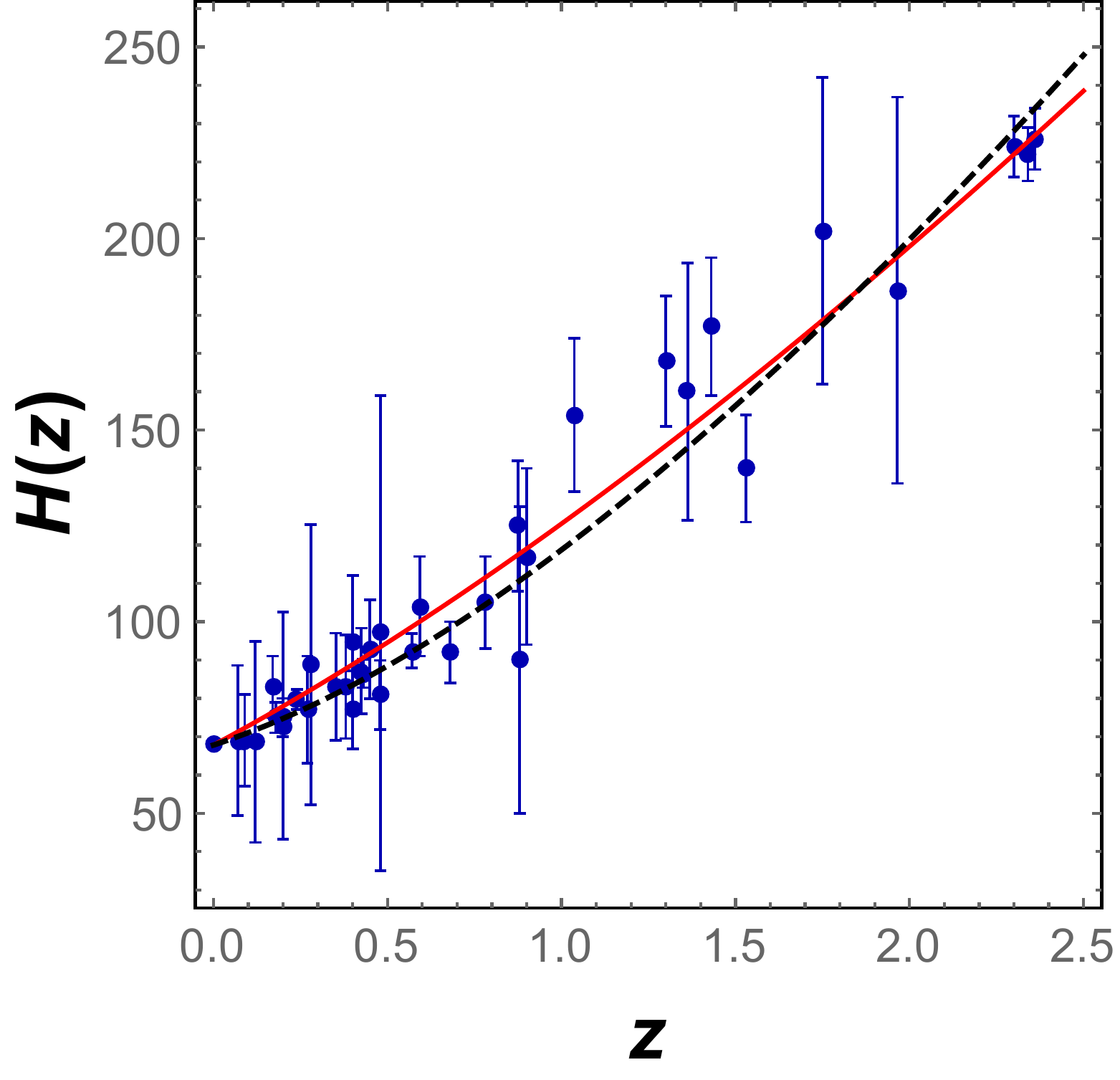} & %
\includegraphics[width=2in]{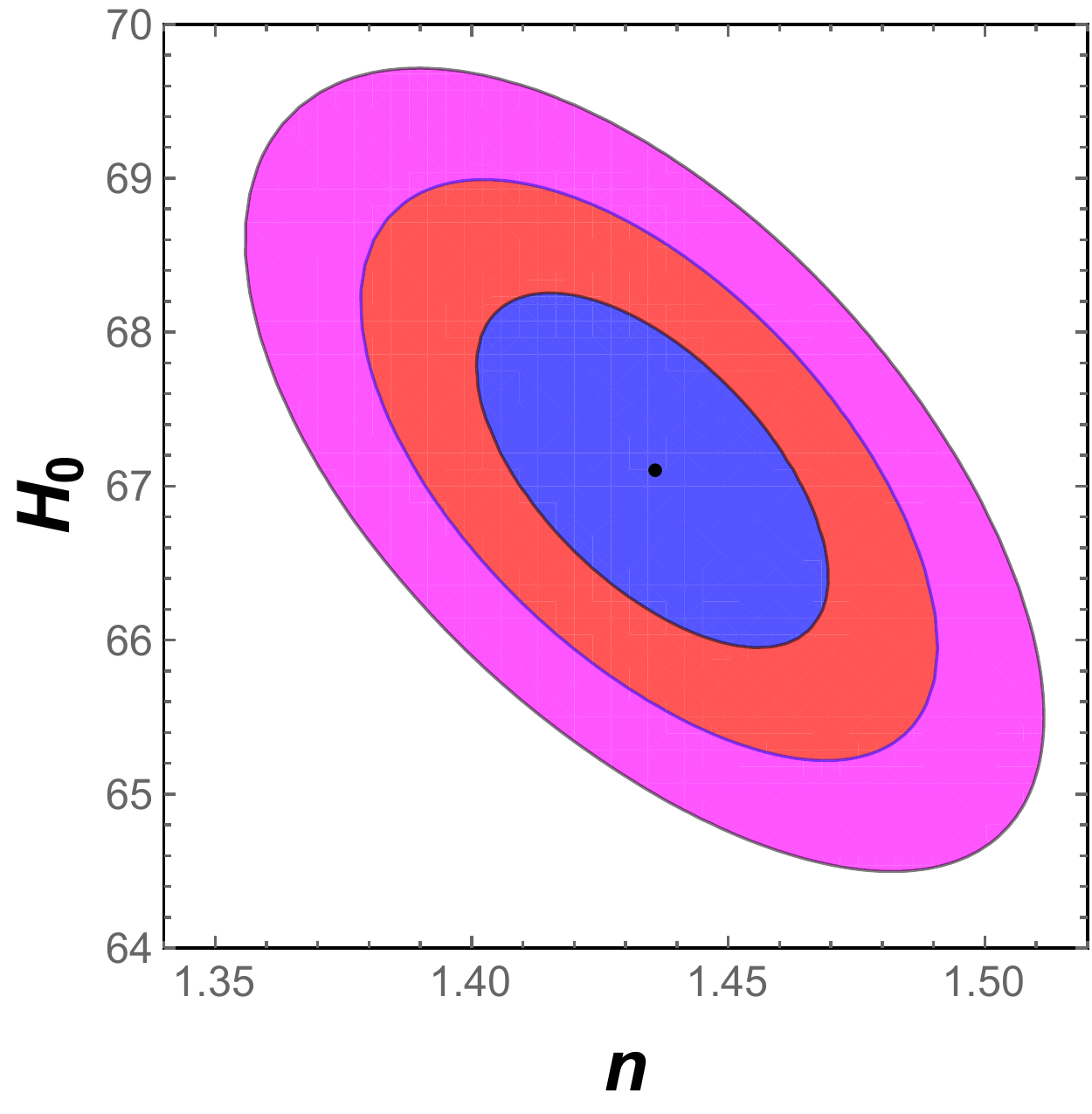} &  \\ 
\mbox (a) & \mbox (b) & 
\end{array}%
$ }
\end{center}
\caption{(a) The error bar plot of $H(z)$ vs. redshift `$z$' and (b) The
contour plot in $n-H_{0}$ plane with the observational Hubble dataset 
\protect\cite{Hzdata}}
\label{fig:12}
\end{figure}

To find a constrained value of model parameter $n$, we have used the same
method of minimizing Chi square value with an updated $36$ points of
observational Hubble dataset (OHD) as used in Ref. \cite{Hzdata}. The above
FIG.\ref{fig:12} demonstrates the error bar plot of $36$ points of OHD
fitted with the $\Lambda $CDM model and our obtained model together with the
constrained values of $n$ and $H_{0}$ as a contour plot in $n$-$H_{0}$ plane
at $1$-$\sigma $, $2$-$\sigma $, $3$-$\sigma $ level. The constrained values
of the model parameter $n$ is found to be $n=1.435$ and $H_{0}=67.102$ with
minimum Chi square value $\chi _{\min }^{2}=19.075$.

\section{Concluding remarks}

\qquad In this work, we have revisited a cosmological model based on General
Theory of Relativity in the FLRW space-time. In view of the observed current
cosmic acceleration and to achieve an exact solution of the cosmological
field equations, we have endorsed a simple parametrization of Hubble
parameter $H$ that turns a time-dependent deceleration parameter $q$ and an
exponential type evolution of the scale factor. The behavior of the
geometrical parameters $a(t)$, $H(t)$, $q(t)$ and physical parameters such
as energy densities of radiation, matter and the dark energy (cosmological
constant here) for the obtained model have been analyzed in detail.
Comprehensive observations for the given cosmological model have been
recorded as follows.\newline

\begin{itemize}
\item A geometrical parametrization of Hubble parameter $H$ used by J. P.
Singh \cite{JPS} and Banerjee et al. \cite{BAN}, have been considered which
leads to a time-dependent deceleration parameter $q$ and can explain the
current acceleration of the universe $(q<0)$ with a prior deceleration $%
(q>0) $ in the past. It has been observed that the universe does not follow
the standard big bang scenario, rather it starts with a finite volume
together with a finite velocity and finite acceleration and is a distinctive
feature against the standard model.

\item The model with a deceleration parameter is time-dependent having
signature flipping behavior with evolution, \textit{i.e.} initially for
small $t$, $q$ is positive and at later stages of the evolution, for large $%
t $ (as $t\rightarrow \infty $), $q$ approach to $-1$. So, the feature of an
early deceleration to late acceleration of the model is suitable for
structure formation in the early stage of evolution and accelerated
expansion in the later stage of the evolution.

\item The dynamics of the obtained model have been discussed in detail in
section III. By considering the cosmological constant as a candidate of dark
energy for which EoS $w_{de}=-1$, we have discussed the evolution of
physical parameters in different eras of the universe. Also from the
expressions of energy densities of radiation $(\rho _{r})$ and dark energy $%
(\rho _{de})$, it has been observed that the positivity condition of energy
densities hold good for the mentioned choices of $n,c,\alpha $ only for flat
and closed geometry of the universe. The chosen numerical values of model
parameters $n,c,\alpha $ fail to satisfy the positivity condition of energy
densities ($\rho _{r}$) and ($\rho _{de}$) for open universe. The profile of
energy densities of radiation and dark energy have been depicted in FIG.\ref%
{fig:3} $-$ \ref{fig:6} respectively for some specific values of the model
parameter for flat and closed geometry by fixing model parameter $\alpha $
and varying $n$ and vice-versa respectively. The profile of varying
radiation temperature \textit{w.r.t} time $t$ in the very early universe has
been highlighted in figures for flat and closed geometries of the universe
with specific values of model parameters. Radiation temperature unfolds like
radiation energy density \textit{i.e.} high temperature initially, then
falls as time goes by and eventually approaches to a constant value in late
time (see FIG.\ref{fig:7} \& \ref{fig:8}).

\item We have examined the matter dominated era for which the dust pressure
reduces to zero. In order to understand the formation of structures in the
universe and late time behavior of the cosmological parameters, we have
established the time-redshift $(t-z)$ relation and expressed geometrical
parameters ($H$ \& $q$) and physical parameters ($\rho _{m}$ \& $\rho _{de}$%
) in terms of redshift. After converting the cosmological parameters in
terms of redshift $z$, it has been noticed that all the parameters are
concerned with $n$ only. The evolutionary profile of energy densities of
matter and dark energy have been represented in FIG.\ref{fig:9} \& \ref%
{fig:10} for flat and closed geometry respectively with variable $n$. The
infringement of the positivity condition of $\rho _{de}$ has been seen for
both flat and closed geometry of the universe for $n=1.63$. The graphical
observation of $\rho _{de}$ caution the violation of its positivity
criterion for the range $n>1.49$. This observation have been evidenced in
the subsequent section using an updated $H(z)$ observational dataset of $36$
points which is an advancement of the work done in \cite{RIT}.

\item In the last section, we have investigated the phase transition of the
deceleration parameter. Recent cosmological observations indicate that the
universe experiences a cosmic speed up at a late time which means that the
universe must have passed through a slower expansion phase in the past. The
cosmic transition from deceleration to acceleration or the phase transition
may be treated as a necessary phenomenon while describing the dynamics of
the universe. Also, recent observation favors the transition at redshift $%
z\approx 0.7$. For our considered parametrization of $H$, we have chosen the
model parameter $n=1.43$ carefully, so that phase transition occurred around 
$z\approx 0.7$. The plot of $q$ vs. $z$ exhibits cosmic deceleration for
high redshift $z$, acceleration for low redshift $z$ and eventually $%
q\rightarrow -1$ as $z\rightarrow -1$. We have plotted the graph of $E(z)=%
\frac{H(z)}{H_{0}}$ \textit{w.r.t} $z$ for the choice of model parameter $%
n=1.43$ (see FIG.\ref{fig:11}).

\item Also, in order to justify the choice of our model parameter $n=1.43$
for which the present model exhibits the phase transition scenario, we have
constrained the value of model parameter $n$ with the help of data analysis.
We have used the same method of minimizing Chi square value with an updated $%
36$ points of observational Hubble dataset (OHD) as used in Ref. \cite%
{Hzdata}. The error bar plot of $36$ points of OHD and the contour plot in $%
n $-$H_{0}$ plane at $1$-$\sigma $, $2$-$\sigma $, $3$-$\sigma $ level are
shown in FIG.\ref{fig:12} (a) \& (b). The presented model has a nice fit to
the OHD. The constrained values of the model parameter $n$ is found to be $%
n=1.435$ and $H_{0}=67.102$ with minimum Chi square value $\chi _{\min
}^{2}=19.075$. In the Ref. \cite{RIT}, the authors have used $29$ points of
OHD and found $n=1.410$, $H_{0}=66.976$ with $\chi _{\min }^{2}=24.579$ for
the same parametrization of $H$ considered here. The constrained values of $%
n $ and $H_{0}$ are preferred as suggested in \cite{Zhang-2, Plank2}.

\item The model presented here is an attempt to address the late-time cosmic
acceleration in FLRW background with a parametrization of $H$ and can also
be extended to the anisotropic and inhomogeneous background. Moreover, some
more issues like big bang nucleosynthesis, structure formation, inflation
can also be discussed in this scenario. Recently, a robust method based on
the redshift dependence of Alcock-Paczynski test is developed in \cite{Li}
to measure the expansion history of the universe that uses the isotropy of
the galaxy density gradient field to provide more tighter constraints on
cosmological parameters with high precision and are studied in a series of
papers by Li et al. \cite{Li-1, Li-2, Li-3, Zhang-1}. The model presented
here and other models with such parametrization could be studied in the same
line to get better and more tighter constraints on the model parameters
using some more datasets and is defer to our future works.
\end{itemize}

\end{document}